\documentclass[11pt,a4paper]{article}
\usepackage{lmodern}
\usepackage[latin9]{inputenc}
\usepackage{color}
\usepackage{amsmath}
\usepackage{amssymb}

\makeatletter

\pdfpageheight\paperheight
\pdfpagewidth\paperwidth

\pdfoutput=0 

\usepackage{jheppub}

\usepackage{etoolbox}
    \makeatletter
    \patchcmd{\maketitle}{\@fpheader}{}{}{}
    \makeatother

\usepackage{color}

\usepackage{amsfonts}

\usepackage{bbm}

\title{\boldmath Integrable systems and the boundary dynamics of higher spin gravity on AdS$_3$}

\author[a,b]{Emilio Ojeda,}
\author[a]{and Alfredo P\'{e}rez}

\affiliation[a]{Centro de Estudios Cient\'{i}ficos (CECs), Avenida Arturo Prat 514, Valdivia,
Chile.}
\affiliation[b]{Departamento de F\'isica, Universidad de Concepci\'on, Casilla 160-C, Concepci\'on, Chile.}

\emailAdd{ojeda@cecs.cl}
\emailAdd{aperez@cecs.cl}

\preprint{CECS-PHY-20/02}%

\abstract{We introduce a new set of boundary conditions for three-dimensional
higher spin gravity with gauge group $SL(3,\mathbb{R})\times SL(3,\mathbb{R})$,
where its dynamics at the boundary is described by the members of
the modified Boussinesq integrable hierarchy. In the asymptotic region
the gauge fields are written in the diagonal gauge, where the excitations
go along the generators of the Cartan subalgebra of $sl(3,\mathbb{R})\oplus sl(3,\mathbb{R})$.
We show that the entire integrable structure of the modified Boussinesq hierarchy, i.e.,
the phase space, the Poisson brackets and the infinite number of commuting conserved charges, are
obtained from the asymptotic structure
of the higher spin theory. Furthermore, its known relation
with the Boussinesq hierarchy is inherited from our analysis once
the asymptotic conditions are re-expressed in the highest weight gauge. Hence, the Miura map is recovered from a purely geometric construction
in the bulk. Black holes that fit within our boundary conditions,
the Hamiltonian reduction at the boundary, and the generalization
to higher spin gravity with gauge group $SL(N,\mathbb{R})\times SL(N,\mathbb{R})$
are also discussed.
}

\makeatother

\begin{document}
\maketitle

\newpage

\section{Introduction\label{sec:1 Introduction}}

The asymptotic structure of spacetime plays a fundamental role in
the description of General Relativity in three dimensions. This theory
does not possess local propagating degrees of freedom, and consequently,
its dynamics is completely dictated by the choice of boundary conditions.
In the case of a negative cosmological constant, it is standard practice
the use of the ones of Brown and Henneaux \cite{Brown:1986nw}, whose
asymptotic symmetries are spanned by two copies of the Virasoro algebra
with central charge $c=3l/2G$. However, this choice is not unique.
There are other possible sensible boundary conditions that can be
consistently used, with different physical consequences \cite{Compere:2013bya,Troessaert:2013fma,Avery:2013dja,Afshar:2016wfy,Perez:2016vqo,Grumiller:2016pqb,Ojeda:2019xih,Perez:2020klz}.
In particular, the ones proposed in ref. \cite{Perez:2016vqo} relate
the dynamics of the gravitational field with the two-dimensional Korteweg-de
Vries (KdV) integrable hierarchy. Therefore, the asymptotic symmetries
are spanned by the infinite set of commuting KdV charges, which are
composite in terms of the Virasoro generators. This framework allows
for example to study Generalized Gibbs Ensembles of two-dimensional
conformal field theories in terms of a gravitational dual \cite{deBoer:2016bov,Perez:2016vqo},
as well as constructing black hole solutions carrying non-trivial
KdV charges \cite{Dymarsky:2020tjh}.

Different extensions of this relation between a three-dimensional
gravitational theory in the bulk with specific boundary conditions,
and an integrable system at the boundary, were also studied in ref.
\cite{Fuentealba:2017omf} for General Relativity with a vanishing
cosmological constant, in ref. \cite{Melnikov:2018fhb} for gravity
on AdS coupled to two $U(1)$ Chern-Simons fields, and in ref. \cite{Ojeda:2019xih}
for pure gravity on AdS in connection with the Gardner hierarchy.

It is then natural to explore the possibility of generalize these
results to the case of General Relativity coupled to higher spin fields
\cite{Blencowe:1988gj,Bergshoeff:1989ns,Vasiliev:1995dn}. Indeed,
in ref. \cite{Compere:2013gja} a connection between spin-3 gravity
and a ``Boussinesq equation in the light-cone'' was pointed out.
In refs. \cite{Gutperle:2014aja,Beccaria:2015iwa}, some particular
boundary conditions for higher spin gravity with gauge group $SL(N,\mathbb{R})\times SL(N,\mathbb{R})$
were associated to a generalized KdV hierarchy, and some particular
cases, including the Boussinesq equation, were explicitly worked out.
On the other hand, in ref. \cite{Perez:2016vqo} a very precise link
with the Boussinesq hierarchy was described for spin-3 gravity, as
well as for its extension including fields with arbitrary higher spins.
The analysis was based on boundary conditions defined in the highest
weight gauge with a particular choice of chemical potentials, generalizing
the results obtained for the KdV hierarchy in pure gravity.

The Boussinesq equation was first introduced by Joseph Boussinesq
in 1872 in the context of the study of the propagation of one-dimensional
long waves in shallow water, moving in both directions\textcolor{red}{{}
}\cite{Boussinesq1872}. Long after, in 1974, it was realized that
the equation was integrable, and that belongs to a hierarchy of differential
equations \cite{Zakharov:1974}. Most of the important properties
of the Boussinesq equation, including its infinite set of commuting
conserved charges, can be easily derived when a potential equation,
called ``modified Boussinesq'' (mBoussinesq), is introduced (see
e.g \cite{Kaup:1975}). Both equations are then related by an appropriate
generalization of the Miura transformation \cite{Hirota:1977,Fordy:1981,Drinfeld:1984qv}.

One of the purposes of our work is to show that the asymptotic dynamics
of spin-3 gravity on AdS$_{3}$ endowed with a special class of boundary
conditions, is precisely described by the members of the mBoussinesq
hierarchy. In this framework, the gauge fields are defined in the
``diagonal gauge,'' where the excitations go along the generators
of the Cartan subalgebra of $sl(3,\mathbb{R})\oplus sl(3,\mathbb{R})$
\cite{Grumiller:2016kcp}. The link with the integrable system is
then obtained by choosing the chemical potentials as precise functionals
of the dynamical fields, in a way consistent with the action principle.
Hence, the entire integrable structure of the mBoussinesq hierarchy,
i.e., the phase space, the fundamental Poisson brackets given by two
independent $\hat{u}\left(1\right)$ current algebras, and the infinite
set of Hamiltonians in involution, are obtained from the asymptotic
structure of the higher spin theory in the bulk. Furthermore, the
relation with the Boussinesq hierarchy previously found in ref. \cite{Perez:2016vqo}
is inherited from our analysis once the asymptotic conditions are
re-expressed in the highest weight gauge along the lines of ref. \cite{Grumiller:2016kcp}.
Thus, the Miura map is recovered from a purely geometric construction
in the higher spin theory. Black hole solutions that fit within our
boundary conditions, the Hamiltonian reduction at the boundary, and
the generalization to higher spin gravity with gauge group $SL(N,\mathbb{R})\times SL(N,\mathbb{R})$
are also discussed.

The plan of the paper is the following. In the next section we briefly
review the bi-Hamiltonian formulation of the mBoussinesq hierchy and
its relation with the Boussinesq one. In section \ref{sec:2 mBq hierarchy from gravity on AdS}
we propose a new set of boundary conditions for higher spin gravity
on AdS$_{3}$ with gauge group $SL(3,\mathbb{R})\times SL(3,\mathbb{R})$,
such that the asymptotic dynamics is precisely described by a particular
member of the mBoussinesq hierarchy. All the relevant properties of
the integrable system, including its infinite number of commuting
conserved charges, are derived from the theory in the bulk. Section
\ref{sec:3 Black holes} is devoted to the study of black hole configurations
that fit within our boundary conditions. It is shown that they are
generically described by static solutions of the corresponding member
of the mBoussinesq hierarchy. The regularity conditions that come
from requiring trivial holonomies around the thermal cycle, as well
as their thermodynamic properties are also analyzed. In section \ref{sec:3 Reduction-of-the-Chern-Simons}
we perform the Hamiltonian reduction of the Chern-Simons action describing
spin-3 gravity with the boundary conditions associated to the mBoussinesq
hierarchy. It is shown that the boundary dynamics, and in particular
the non-trivial interacting terms in the reduced action are completely
determined by the Hamiltonians of the hierarchy. In section \ref{sec:Some-extensions:-Generalized}
we discuss two possible extensions of our previous results. We first
describe a natural generalization of the boundary conditions that
allows to study Generalized Gibbs Ensembles by considering a general
Hamiltonian consisting of a linear combination of all the charges
in involution of the hierarchy. We also extend our results to the
case of three-dimensional higher spin gravity with gauge group $SL(N,\mathbb{R})\times SL(N,\mathbb{R})$,
where the associated hierarchy correspond to the $N$-th ``Gelfand-Dickey
hierarchy.'' Some final remarks and further possible extensions are
briefly addressed in section \ref{sec:5 Final-remarks}. Five appendices
are included. In appendix \ref{sec:Second-Hamiltonian-structure}
the operator that defines the second Poisson structure of the mBoussinesq
hierarchy is exhibited. In appendix \ref{sec:8 Gelfand-Dikii-polynomials-and}
the first Hamiltonians and Gelfand-Dickey polynomials of the mBoussinesq
hierarchy are displayed. Appendix \ref{sec:7 Boussinesq-and-modified}
is devoted to a brief review of the Boussinesq hierarchy. In appendix
\ref{sec:6 SL(3,R)-in-the}, the fundamental matrix representation
of $SL(2,\mathbb{R})$ within $SL(N,\mathbb{R})$, the principal embedding
is presented. Finally, in appendix \ref{sec:Wess-Zumino-term} it
is shown that for our boundary conditions the Wess-Zumino term in
the Hamiltonian reduction in section \ref{sec:3 Reduction-of-the-Chern-Simons}
vanishes.

\section{Review of the modified Boussinesq hierarchy\label{sec:Review-of-the}}

The first member of the mBoussinesq hierarchy is given by the following
set of differential equations

\begin{eqnarray}
\dot{\mathcal{J}} & = & \lambda_{1}\mathcal{J}^{\prime}-\lambda_{2}\left(2\left(\mathcal{J}\mathcal{U}\right)^{\prime}+\mathcal{U}^{\prime\prime}\right),\nonumber \\
\dot{{\cal U}} & = & \lambda_{1}\mathcal{U}^{\prime}+\lambda_{2}\left(\mathcal{U}^{2\prime}-\mathcal{J}^{2\prime}+\mathcal{J}^{\prime\prime}\right).\label{eq:mBsq}
\end{eqnarray}
Here dots and primes denote derivatives with respect to the time $t$
and the angle $\phi$ respectively, and $\lambda_{1}$, $\lambda_{2}$
are arbitrary constants associated to the two different flows of the
hierarchy \cite{Fordy:1981}. The case with $\lambda_{1}=0$ and $\lambda_{2}=1$
is known as the mBoussinesq equation, while the case with $\lambda_{1}=1$
and $\lambda_{2}=0$ describes two independent chiral fields.

The dynamics of the above equations may be described using the Hamiltonian
formalism. If the Poisson brackets of two arbitrary functional $F$
and $G$ is given by

\begin{equation}
\left\{ F,G\right\} =\frac{4\pi}{\hat{\kappa}}\int d\phi\left(\frac{\delta F}{\delta\mathcal{J}}\partial_{\phi}\frac{\delta G}{\delta\mathcal{J}}+\frac{\delta F}{\delta\mathcal{U}}\partial_{\phi}\frac{\delta G}{\delta\mathcal{U}}\right),\label{eq:Poissonbrack}
\end{equation}
together with the Hamiltonian

\begin{equation}
H_{\left(1\right)}=\frac{\hat{\kappa}}{4\pi}\int d\phi\left\{ \frac{\lambda_{1}}{2}\left(\mathcal{J}^{2}+\mathcal{U}^{2}\right)+\lambda_{2}\left(\frac{1}{3}\mathcal{U}^{3}-\mathcal{J}^{2}\mathcal{U}-\mathcal{J}\mathcal{U}^{\prime}\right)\right\} ,\label{eq:Ham1}
\end{equation}
then eq. \eqref{eq:mBsq} can be rewritten as
\[
\dot{\mathcal{J}}=\left\{ \mathcal{J},H_{\left(1\right)}\right\} \,,\qquad\qquad\dot{{\cal U}}=\left\{ {\cal U},H_{\left(1\right)}\right\} .
\]
Note that the arbitrary constant $\hat{\kappa}$ does not appear in
the differential equations \eqref{eq:mBsq}, however it is useful
to introduce it in \eqref{eq:Poissonbrack} and \eqref{eq:Ham1} for
later convenience.

Alternatively, if we define the operator 
\begin{equation}
\mathcal{D}:=\frac{4\pi}{\hat{\kappa}}\left(\begin{array}{cc}
\partial_{\phi} & 0\\
0 & \partial_{\phi}
\end{array}\right),\label{eq:Symp1}
\end{equation}
the equations in \eqref{eq:mBsq} can be re-written in vector form
as follows

\[
\left(\begin{array}{c}
\dot{\mathcal{J}}\\
\dot{{\cal U}}
\end{array}\right)=\mathcal{D}\left(\begin{array}{c}
\frac{\delta H_{\left(1\right)}}{\delta\mathcal{J}}\\
\frac{\delta H_{\left(1\right)}}{\delta\mathcal{U}}
\end{array}\right).
\]
The operator $\mathcal{D}$ in \eqref{eq:Symp1} defines the symplectic
structure in eq. \eqref{eq:Poissonbrack}.

It is worth to emphasize that one of the key points in the relation
of this integrable system with higher spin gravity comes from the
fact that, according to eq. \eqref{eq:Poissonbrack}, the fundamental
Poisson brackets are described by two independent $\hat{u}\left(1\right)$
current algebras

\begin{eqnarray}
\left\{ \mathcal{J}\left(\phi\right),\mathcal{J}\left(\phi^{\prime}\right)\right\}  & = & \frac{4\pi}{\hat{\kappa}}\partial_{\phi}\delta\left(\phi-\phi^{\prime}\right),\nonumber \\
\left\{ \mathcal{U}\left(\phi\right),\mathcal{U}\left(\phi^{\prime}\right)\right\}  & = & \frac{4\pi}{\hat{\kappa}}\partial_{\phi}\delta\left(\phi-\phi^{\prime}\right).\label{eq:FUndbrack}
\end{eqnarray}
As we will show below, once appropriate boundary conditions are imposed,
this Poisson bracket algebra is obtained from the Dirac brackets in
the higher spin theory.

The integrability of \eqref{eq:mBsq} and the existence of a hierarchy
of equations, rely on the fact that this system is actually bi-Hamiltonian.
Indeed, there exists an alternative symplectic structure characterized
by the following non-local operator

\begin{equation}
\mathcal{D}_{\left(2\right)}=\mathcal{D}M^{\dagger}\mathcal{O}M\mathcal{D}.\label{eq:secondPstr}
\end{equation}
Here,

\begin{equation}
M=\left(\begin{array}{cc}
{\cal J}+\partial_{\phi}\qquad & \mathcal{U}\\
-2{\cal J}\mathcal{U}-\frac{1}{2}\mathcal{U}\partial_{\phi}-\frac{3}{2}{\cal U}^{\prime}\qquad & \mathcal{U}^{2}-{\cal J}^{2}-\frac{1}{2}{\cal J}^{\prime}-\frac{3}{2}{\cal J}\partial_{\phi}-\frac{1}{2}\partial_{\phi}^{2}
\end{array}\right),\label{eq:M}
\end{equation}
and $M^{\dagger}$ is the formal adjoint of $M$ (see e.g. \cite{Mathieu:1991}).
The operator $\mathcal{O}$ that contains the non-local terms is given
by
\[
\mathcal{O}=\frac{2\hat{\kappa}}{\pi}\left(\begin{array}{cc}
0 & \partial_{\phi}^{-1}\\
\partial_{\phi}^{-1} & 0
\end{array}\right).
\]
 Consequently, the Poisson bracket of two arbitrary functionals $F$
and $G$ associated with the operator $\mathcal{D}_{\left(2\right)}$
is

\begin{equation}
\left\{ F,G\right\} _{2}=\int d\phi\left(\begin{array}{cc}
\frac{\delta F}{\delta\mathcal{J}} & \frac{\delta F}{\delta\mathcal{U}}\end{array}\right)\mathcal{D}_{(2)}\left(\begin{array}{c}
\frac{\delta G}{\delta\mathcal{J}}\\
\frac{\delta G}{\delta\mathcal{U}}
\end{array}\right).\label{eq:Poisson2}
\end{equation}
The explicit components of $\mathcal{D}_{\left(2\right)}$ are exhibited
in appendix \ref{sec:Second-Hamiltonian-structure}.

Equations \eqref{eq:mBsq} can then be recovered using the Poisson
bracket \eqref{eq:Poisson2}, together with the Hamiltonian\footnote{The coefficients $\lambda_{1}$ and $\lambda_{2}$ in eq. \eqref{eq:mBsq}
are determined by the integration constants obtained by the action
of $\mathcal{D}_{\left(2\right)}$. In the case of higher members
of the hierarchy, the subsequent integration constants may be consistently
set to zero as they contribute nothing new (see e.g \cite{Fordy:1981}).}

\[
H_{\left(0\right)}=\frac{\hat{\kappa}}{4\pi}\int d\phi\left(\lambda_{1}\mathcal{J}+\lambda_{2}\mathcal{U}\right).
\]
This system possess an infinite number of conserved charges in involution
that can be constructed from the following recursion relation
\begin{equation}
R_{\left(n+1\right)}=\mathcal{D}^{-1}\mathcal{D}_{\left(2\right)}R_{\left(n\right)},\label{eq:rec}
\end{equation}
where
\begin{equation}
R_{\left(n\right)}=\left(\begin{array}{c}
\frac{\delta H_{\left(n\right)}}{\delta\mathcal{J}}\\
\frac{\delta H_{\left(n\right)}}{\delta\mathcal{U}}
\end{array}\right),\label{eq:GelfandDikii}
\end{equation}
are the Gelfand-Dickey polynomials associated to the hierarchy. The
conserved quantities $H_{\left(n\right)}$, with $n$ being a nonnegative
integer, are generically decomposed into two flows proportional to
the constants $\lambda_{1}$ and $\lambda_{2}$ respectively
\begin{equation}
H_{\left(n\right)}=\sum_{I=1}^{2}\lambda_{I}H_{\left(n\right)}^{I}.\label{eq: decomp}
\end{equation}
Then one can prove that the $H_{\left(k\right)}^{I}$ are in involution
with both Poisson brackets, i.e.,
\[
\left\{ H_{\left(n\right)}^{I},H_{\left(m\right)}^{J}\right\} =\left\{ H_{\left(n\right)}^{I},H_{\left(m\right)}^{J}\right\} _{2}=0.
\]
Furthermore, if we one uses the conserved quantities $H_{\left(k\right)}^{I}$
as new Hamiltonians, it is then possible to define a hierarchy of
integrable equations labelled by the nonnegative integer $k$ of the
form

\begin{eqnarray}
\dot{\mathcal{J}} & = & \left\{ \mathcal{J},H_{\left(k\right)}\right\} =\left\{ \mathcal{J},H_{\left(k-1\right)}\right\} _{2},\nonumber \\
\dot{{\cal U}} & = & \left\{ {\cal U},H_{\left(k\right)}\right\} =\left\{ {\cal U},H_{\left(k-1\right)}\right\} _{2}.\label{eq:hierarchy}
\end{eqnarray}

The equations associated to each flow, labelled by the index $I=1,2$,
have different scaling properties. Under a Lifshitz scaling transformation
with dynamical exponent $z$

\[
t\rightarrow\varepsilon^{z}t\,,\quad\quad\phi\rightarrow\varepsilon\phi\,,\quad\quad\mathcal{J}\rightarrow\varepsilon^{-1}\mathcal{J}\,,\quad\quad\mathcal{U}\rightarrow\varepsilon^{-1}\mathcal{U},
\]
the flow with $I=1$ is invariant for $z=3k-2$, while the flow with
$I=2$ is invariant for $z=3k-1$.

As explained in the introduction, the mBoussinesq equation is a ``potential
equation'' for the Boussinesq one. Indeed, if $\mathcal{U}$ and
$\mathcal{J}$ obey the mBoussinesq equation, then the fields $\mathcal{L}$
and $\mathcal{W}$ defined by the Miura transformation as
\begin{align}
\mathcal{L} & =\frac{1}{2}{\cal J}^{2}+\frac{1}{2}\mathcal{U}^{2}+{\cal J}^{\prime},\nonumber \\
\mathcal{W} & =\frac{1}{3}\mathcal{U}^{3}-{\cal J}^{2}\mathcal{U}-\frac{1}{2}\mathcal{U}{\cal J}^{\prime}-\frac{3}{2}{\cal J}{\cal U}^{\prime}-\frac{1}{2}{\cal U}^{\prime\prime},\label{eq:Miura}
\end{align}
obey the Boussinesq equation given by
\begin{eqnarray}
\dot{\mathcal{L}} & = & 2\mathcal{W}^{\prime},\nonumber \\
\dot{\mathcal{W}} & = & 2\mathcal{L}^{2\prime}-\frac{1}{2}\mathcal{L}^{\prime\prime\prime}.\label{eq:Bsqeq}
\end{eqnarray}

The entire Boussinesq hierarchy, including the infinite set of charges
in involution, is obtained from the mBoussinesq one by using the Miura
transformation (see appendix \ref{sec:7 Boussinesq-and-modified}\textcolor{red}{{}
}for more details on the Boussinesq hierarchy). It is worth noting
that the Miura transformation \eqref{eq:Miura} coincides the twisted
Sugawara construction of the stress tensor and a spin-3 current in
terms of two independent $U\left(1\right)$ currents in a two-dimensional
CFT. Hence, using the Poisson brackets \eqref{eq:FUndbrack} and the
Miura map \eqref{eq:Miura}, one can show that the (first) Poisson
brackets for $\mathcal{L}$ and $\mathcal{W}$ are precisely given
by the classical $W_{3}$ algebra.

\section{Modified Boussinesq hierarchy from spin-3 gravity on AdS$_{3}$\label{sec:2 mBq hierarchy from gravity on AdS}}

\subsection{Chern-Simons formulation of spin-3 gravity on AdS$_{3}$}

Higher spin gravity in 3D has the very special property that, in contrast
with their higher dimensional counterparts \cite{Fradkin:1987ks,Vasiliev:1990en,Vasiliev:2003ev},
its spectrum can be consistently truncated to a finite number of higher
spin fields \cite{Blencowe:1988gj,Bergshoeff:1989ns,Henneaux:2010xg,Campoleoni:2010zq}.
One of the simplest cases corresponds to a spin-two field non-minimally
coupled to a spin-three field, that may be described by a Chern-Simons
action for the gauge group $SL\left(3,\mathbb{R}\right)\times SL\left(3,\mathbb{R}\right)$,

\begin{equation}
I=I_{CS}\left[A^{+}\right]-I_{CS}\left[A^{-}\right],\label{eq:ICS}
\end{equation}
where

\begin{equation}
I_{CS}[A]=\frac{\kappa_{3}}{4\pi}\int_{\mathcal{M}}\text{tr}\left(AdA+\frac{2}{3}A^{3}\right).\label{eq:Chern-Simons Action}
\end{equation}
Here, the level is given by \emph{$\kappa_{3}=\kappa/4=l/16G$}, where
$l$ and $G$ correspond to the AdS radius and the three-dimensional
Newton constant respectively, and the trace is in the fundamental
representation of the $sl\left(3,\mathbb{R}\right)$ algebra in the
principal embedding (see appendix \ref{sec:6 SL(3,R)-in-the}). The
field equations are then given by the vanishing of the field strength
\begin{equation}
F^{\pm}=dA^{\pm}+A^{\pm2}=0.\label{eq:Field Strength}
\end{equation}
The metric and the spin-three field are reconstructed in terms of
a generalized dreibein $e:=\frac{l}{2}\left(A^{+}-A^{-}\right)$ according
to

\[
g_{\mu\nu}=\frac{1}{2}\textrm{tr}\left(e_{\mu}e_{\nu}\right)\,,\qquad\qquad\varphi_{\mu\nu\rho}=\frac{1}{3!}\textrm{tr}\left(e_{\left(\mu\right.}e_{\nu}e_{\left.\rho\right)}\right).
\]

\subsection{Asymptotic behavior of the fields. Diagonal gauge\label{subsec:2.1 Asymptotic behavior of the fields. Diagonal gauge}}

Following refs. \cite{Coussaert:1995zp,Henneaux:2010xg,Campoleoni:2010zq},
it is convenient to perform the analysis of the asymptotic symmetries
of spin-three gravity in terms of an auxiliary connection depending
only on $t$ and $\phi$. For simplicity, and without loss of generality,
hereafter we will consider only the ``plus copy,'' and hence the
superscript ``$+$'' will be omitted. The gauge field $A$ is then
written as

\begin{equation}
A=b^{-1}\left(d+a\right)b,\label{eq:A Connections}
\end{equation}
where $a=a_{t}dt+a_{\phi}d\phi$ is the auxiliary connection, and
$b=b\left(r\right)$ is a gauge group element which captures the whole
the radial dependence of the gauge connection. The asymptotic analysis
will be insensitive to the precise form of $b\left(r\right)$.

We will consider asymptotic conditions in the ``diagonal gauge,''
i.e., where all the permissible excitations in the auxiliary connection
go along the generators of the Cartan subalgebra of $sl\left(3,\mathbb{R}\right)$
\cite{Grumiller:2016kcp}. Then, it takes the form

\begin{equation}
a=\left({\cal J}d\phi+\zeta dt\right)L_{0}+\frac{\sqrt{3}}{2}\left(\mathcal{U}d\phi+\zeta_{\mathcal{U}}dt\right)W_{0}.\label{eq:Auxiliary Connections}
\end{equation}
The fields \emph{${\cal J}$} and\emph{ $\mathcal{U}$} belong to
the spatial components of the auxiliary connection, and hence they
are identified as the dynamical fields. On the other hand, \emph{$\zeta$
}and\emph{ $\zeta_{\mathcal{U}}$} are defined along the temporal
components, and therefore they correspond to the boundary values of
the Lagrange multipliers. In ref. \cite{Grumiller:2016kcp} the same
asymptotic form for the auxiliary connection was used, with the replacement
$\mathcal{J}_{\left(3\right)}\rightarrow\frac{\sqrt{3}}{2}\mathcal{U}$
and $\zeta_{\left(3\right)}\rightarrow\frac{\sqrt{3}}{2}\zeta_{\mathcal{U}}$.
However, the boundary conditions will be different. In \cite{Grumiller:2016kcp}
it was assumed that $\zeta$ and $\zeta_{\left(3\right)}$ are kept
fixed at the boundary, while here, as we will show in the next subsection,
they will acquire a precise functional dependence on the dynamical
fields ${\cal J}$, ${\cal U}$ and their spatial derivatives.

\subsection{Boundary conditions for spin-3 gravity and the modified Boussinesq
hierarchy\label{subsec:2.2 Consistency-with-the}}

A fundamental requirement in the study of the asymptotic structure
of spacetime is that the boundary conditions must be compatible with
the action principle. In the canonical formalism one has to add an
appropriate boundary term $B_{\infty}$ to the canonical action in
order to guarantee that the action principle attains an extremum \cite{Regge:1974zd}
\begin{equation}
I_{can}\left[A\right]=-\frac{\kappa}{16\pi}\int dtd^{2}x\epsilon^{ij}\left\langle A_{i}\dot{A_{j}}-A_{t}F_{ij}\right\rangle +B_{\infty}.\label{eq:Action in hamiltonian form}
\end{equation}
Following \cite{Grumiller:2016kcp}, for the action \eqref{eq:Action in hamiltonian form}
and the asymptotic conditions \eqref{eq:A Connections}, \eqref{eq:Auxiliary Connections},
the variation of the boundary term is given by
\begin{equation}
\delta B_{\infty}=-\frac{\kappa}{4\pi}\int dtd\phi\left(\zeta\delta\mathcal{J}+\zeta_{\mathcal{U}}\delta\mathcal{U}\right).\label{eq:Variation of boundary terms}
\end{equation}
In the absence of ingoing or outgoing radiation, as is the case in
three-dimensional higher spin gravity, the boundary term $B_{\infty}$
has to be integrable in a functional sense, i.e., one must be able
to ``take the delta outside'' in \eqref{eq:Variation of boundary terms}.
The precise form in which $\zeta$ and $\zeta_{\mathcal{U}}$ are fixed
at the boundary is what defines the boundary conditions. Thus, following
\cite{Perez:2016vqo}, in order to make contact with the Boussinesq
hierarchy in a way consistent with the action principle, we choose
the Lagrange multipliers as
\begin{equation}
\zeta=\frac{4\pi}{\kappa}\frac{\delta H_{\left(k\right)}}{\delta\mathcal{J}}\,,\qquad\qquad\zeta_{\mathcal{U}}=\frac{4\pi}{\kappa}\frac{\delta H_{\left(k\right)}}{\delta\mathcal{U}},\label{eq:chemical potentials}
\end{equation}
where $H_{\left(k\right)}$ is the Hamiltonian associated to the $k$-th
element of the mBoussinesq hierarchy. With this choice, the boundary
term can be readily integrated, and yields

\begin{equation}
B_{\infty}=-\int dtH_{\left(k\right)}.\label{eq:Boundary terms in terms of H}
\end{equation}
Thus, the Hamiltonian of the gravitational theory in the reduced phase
space precisely matches the one of the integrable system.

On the other hand, the field equations in the higher spin theory given
by the vanishing of the field strength become

\begin{equation}
\dot{\mathcal{J}}=\zeta^{\prime}\,,\qquad\qquad\dot{\mathcal{U}}=\zeta_{\mathcal{U}}^{\prime}\,,\label{eq:Field equations}
\end{equation}
which, by virtue of \eqref{eq:chemical potentials}, precisely coincide
with the differential equations associated to the $k$-th element
of the mBoussinesq hierarchy in eq. \eqref{eq:hierarchy}, provided
the constant $\hat{\kappa}$ in eqs. \eqref{eq:Poissonbrack} and
\eqref{eq:Ham1} is assumed to depend on the cosmological and Newton
constants according to $\hat{\kappa}=\kappa$. Then,

\begin{equation}
\left(\begin{array}{c}
\dot{\mathcal{J}}\\
\dot{{\cal U}}
\end{array}\right)=\mathcal{D}\left(\begin{array}{c}
\frac{\delta H_{\left(k\right)}}{\delta\mathcal{J}}\\
\frac{\delta H_{\left(k\right)}}{\delta\mathcal{U}}
\end{array}\right)=\left(\begin{array}{c}
\left\{ \mathcal{J},H_{\left(k\right)}\right\} \\
\left\{ {\cal U},H_{\left(k\right)}\right\} 
\end{array}\right).\label{eq:EOM}
\end{equation}

\subsection{Asymptotic symmetries and conserved charges\label{subsec:2.3 Asymptotic-symmetries-and}}

The asymptotic symmetries are determined by set of gauge transformations
that preserve the asymptotic form of the gauge connection, with non-vanishing
associated charges. The form of the auxiliary connection in eq.
\eqref{eq:Auxiliary Connections} is preserved by gauge transformations
$\delta a=d\lambda+\left[a,\lambda\right]$, with parameter

\[
\lambda=\eta L_{0}+\frac{\sqrt{3}}{2}\eta_{{\cal U}}W_{0}.
\]
There could be some additional terms in the non-diagonal components
of the gauge parameter $\lambda$, but they are pure gauge in the
sense that there are no generators associated to them, so they can
be consistently set to zero.

The preservation of the angular components of the auxiliary connection
gives the transformation law of the dynamical fields
\begin{equation}
\delta\mathcal{J}=\eta^{\prime}\,,\qquad\qquad\delta{\cal U}=\eta_{{\cal U}}^{\prime},\label{eq:tarnsflawfields}
\end{equation}
while that the preservation of the temporal components provides the
transformation law of the Lagrange multipliers
\begin{equation}
\delta\zeta=\dot{\eta}\,,\qquad\qquad\delta\mathcal{\zeta}_{{\cal U}}=\dot{\eta}_{{\cal U}}.\label{eq:transflachem}
\end{equation}
The variation of the conserved charges can be computed using the Regge-Teitelboim
method \cite{Regge:1974zd}, and they are given by the following surface
integral

\begin{equation}
\delta Q\left[\eta,\eta_{{\cal U}}\right]=\frac{\kappa}{4\pi}\int d\phi\left(\eta\delta\mathcal{J}+\eta_{{\cal U}}\delta{\cal U}\right).\label{eq:Variation of the charges}
\end{equation}
The Dirac brackets of the dynamical fields $\mathcal{J}$ and $\mathcal{U}$
induced by the asymptotic conditions may be obtained from the relation
$\delta_{Y}Q\left[X\right]=\left\{ Q\left[X\right],Q\left[Y\right]\right\} $,
and is given by two independent $\hat{u}\left(1\right)$ current algebras

\begin{eqnarray}
\left\{ \mathcal{J}\left(\phi\right),\mathcal{J}\left(\phi^{\prime}\right)\right\} ^{\star} & = & \frac{4\pi}{\kappa}\partial_{\phi}\delta\left(\phi-\phi^{\prime}\right),\nonumber \\
\left\{ \mathcal{U}\left(\phi\right),\mathcal{U}\left(\phi^{\prime}\right)\right\} ^{\star} & = & \frac{4\pi}{\kappa}\partial_{\phi}\delta\left(\phi-\phi^{\prime}\right),\label{eq:Dirac}
\end{eqnarray}
expression that coincides with the first Poisson bracket of the mBoussinesq
hierarchy given by eq. \eqref{eq:FUndbrack}. Furthermore, the infinite
set of commuting charges of the hierarchy is obtained from the surface
integral \eqref{eq:Variation of the charges} as follows: if we take
into account that due to eq. \eqref{eq:chemical potentials} the Lagrange
multipliers are field dependent, then the consistency with their transformation
law \eqref{eq:transflachem} implies the following differential equation
that must be obeyed by $\eta$ and $\eta_{{\cal U}}$

\[
\left(\begin{array}{c}
\dot{\eta}\left(t,\theta\right)\\
\dot{\eta}_{{\cal U}}\left(t,\theta\right)
\end{array}\right)=\int d\phi\left(\begin{array}{cc}
\frac{\delta^{2}H_{\left(k\right)}}{\delta\mathcal{J}\left(t,\theta\right)\delta\mathcal{J}\left(t,\phi\right)} & \frac{\delta^{2}H_{\left(k\right)}}{\delta{\cal U}\left(t,\theta\right)\delta\mathcal{J}\left(t,\phi\right)}\\
\frac{\delta^{2}H_{\left(k\right)}}{\delta\mathcal{J}\left(t,\theta\right)\delta{\cal U}\left(t,\phi\right)} & \frac{\delta^{2}H_{\left(k\right)}}{\delta{\cal U}\left(t,\theta\right)\delta{\cal U}\left(t,\phi\right)}
\end{array}\right)\mathcal{D}\left(\begin{array}{c}
\eta\\
\eta_{{\cal U}}
\end{array}\right).
\]
By virtue of the integrability of the system, the most general solution
of this equation, under the assumption that $\eta$ and $\eta_{{\cal U}}$
depend locally on $\mathcal{J}$, $\mathcal{U}$ and their spatial
derivatives, is given by

\begin{equation}
\left(\begin{array}{c}
\eta\\
\eta_{{\cal U}}
\end{array}\right)=\frac{4\pi}{\kappa}\sum_{n=0}^{\infty}\alpha_{\left(n\right)}\left(\begin{array}{c}
\frac{\delta H_{\left(n\right)}}{\delta\mathcal{J}}\\
\frac{\delta H_{\left(n\right)}}{\delta\mathcal{U}}
\end{array}\right),\label{eq:etas}
\end{equation}
where the $\alpha_{\left(n\right)}$ are arbitrary constants. Therefore,
replacing the solution \eqref{eq:etas} in \eqref{eq:Variation of the charges}
one can integrate the charge in the functional sense (taking the delta
outside), obtaining

\[
Q=\sum_{n=0}^{\infty}\alpha_{\left(n\right)}H_{\left(n\right)}.
\]
Thus, the conserved charges in the higher spin theory are precisely
given by a linear combination of the Hamiltonians of the mBoussinesq
hierarchy. Indeed, using the transformation law \eqref{eq:tarnsflawfields},
one can show the Hamiltonians $H_{\left(n\right)}^{I}$ are in involution
with respect to the Dirac bracket \eqref{eq:Dirac}

\[
\left\{ H_{\left(m\right)}^{I},H_{\left(n\right)}^{J}\right\} ^{\star}=0,
\]
as expected.

In sum, all the relevant properties of the integrable mBoussinesq
hierarchy described in section \ref{sec:Review-of-the} are derived
from spin-3 gravity endowed with the boundary conditions defined in
eqs. \eqref{eq:A Connections}, \eqref{eq:Auxiliary Connections}
and \eqref{eq:chemical potentials}. Thus, the reduced phase space
of spin-3 gravity and its boundary dynamics are equivalent to the
ones of the mBoussinesq hierarchy. In particular, this provides an
explicit one-to-one map between solutions of the integrable system
at the boundary and solutions of the higher spin gravity theory in
the bulk.

\subsection{Highest weight gauge, Miura map and the Boussinesq hierarchy\label{subsec:Relation-with-the}}

Asymptotic conditions for spin-3 gravity on AdS$_{3}$ in the highest
weight gauge were first given in refs. \cite{Campoleoni:2010zq,Henneaux:2010xg},
where it was shown that the asymptotic symmetries are spanned by two
copies of the classical $W_{3}$ algebra with the Brown-Henneaux central
charge. In what follows we consider the generalization introduced
in refs. \cite{Henneaux:2013dra,Bunster:2014mua}, where the most
general form of the Lagrange multipliers $a_{t}$, compatible with
the $W_{3}$ symmetry, is allowed. This generalization has the important
property that it accommodates black holes carrying non-trivial higher
spin charges.

In this subsection we show that, with a particular gauge transformation,
the auxiliary connection in the diagonal gauge \eqref{eq:Auxiliary Connections}
can be mapped to an auxiliary connection in the highest weight gauge,
such that the Miura transformation in eq. \eqref{eq:Miura}, that
relates the mBoussinesq with the Boussinesq hierarchies, is recovered
from a purely geometric construction in the higher spin theory. The
analysis is very close to the one in ref. \cite{Grumiller:2016kcp},
with the main difference that now the Lagrange multipliers $\zeta$
and $\zeta_{\mathcal{U}}$ depend on the dynamical fields according
to eq. \eqref{eq:chemical potentials}.

The angular components of the auxiliary connection in the highest
weight gauge $\hat{a}$ is assumed to be of the form

\begin{equation}
\hat{a}_{\varphi}=L_{1}-\frac{1}{2}\mathcal{L}L_{-1}-\frac{1}{4\sqrt{3}}\mathcal{W}W_{-2}.\label{eq:Auxiliary connection HWg}
\end{equation}
Following \cite{Henneaux:2013dra,Bunster:2014mua}, the most general
form of $\hat{a}_{t}$ which is compatible with the field equations
is

\begin{align}
\hat{a}_{t} & =\mu L_{1}-\frac{\sqrt{3}}{2}\nu W_{2}-\mu^{\prime}L_{0}+\frac{\sqrt{3}}{2}\nu^{\prime}W_{1}+\frac{1}{2}\left(\mu^{\prime\prime}-\mu\mathcal{L}-2\mathcal{W}\nu\right)L_{-1}\nonumber \\
 & \quad-\frac{\sqrt{3}}{48}\left(4\mathcal{W}\mu-7\mathcal{L}^{\prime}\nu^{\prime}-2\nu\mathcal{L}^{\prime\prime}-8\mathcal{L}\nu^{\prime\prime}+6\mathcal{L}^{2}\nu+\nu^{\prime\prime\prime\prime}\right)W_{-2}\nonumber \\
 & \quad-\frac{\sqrt{3}}{4}\left(\nu^{\prime\prime}-2\mathcal{L}\nu\right)W_{0}+\frac{\sqrt{3}}{12}\left(\nu^{\prime\prime\prime}-2\nu\mathcal{L}^{\prime}-5\mathcal{L}\nu^{\prime}\right)W_{-1}\ .\label{eq:AtHWG}
\end{align}
It is possible to find a gauge group element $g=g^{(1)}g^{(2)}$,
such that the auxiliary connection in the diagonal gauge $a$ is mapped
to the auxiliary connection in the highest weight gauge $\hat{a}$,
by a gauge transformation of the form $\hat{a}=g^{-1}\left(d+a\right)g$,
with
\[
\begin{array}{l}
g^{(1)}=\exp\left[xL_{1}+yW_{1}+zW_{2}\right],\\
g^{(2)}=\exp\left[-\frac{1}{2}\mathcal{J}L_{-1}-\frac{\sqrt{3}}{6}\mathcal{U}W_{-1}+\frac{\sqrt{3}}{12}\left(\mathcal{J}\mathcal{U}+\frac{1}{2}\mathcal{U}^{\prime}\right)\mathrm{W}_{-2}\right].
\end{array}
\]
Here, the functions $x$, $y$, $z$ are restricted to obey the following
differential equations

\[
\begin{array}{l}
x^{\prime}=1+x\,\mathcal{J}+\sqrt{3}y\,\mathcal{U},\\
y^{\prime}=y\,\mathcal{J}+\sqrt{3}x\,\mathcal{U},\\
z^{\prime}=-\frac{1}{2}y+2z\,\mathcal{J}.
\end{array}
\]
The fields $\mathcal{L}$ and $\mathcal{W}$ are then related to the
fields $\mathcal{J}$ and $\mathcal{U}$ precisely by the Miura transformation
\eqref{eq:Miura}
\begin{align}
\mathcal{L} & =\frac{1}{2}{\cal J}^{2}+\frac{1}{2}\mathcal{U}^{2}+{\cal J}^{\prime},\nonumber \\
\mathcal{W} & =\frac{1}{3}\mathcal{U}^{3}-{\cal J}^{2}\mathcal{U}-\frac{1}{2}\mathcal{U}{\cal J}^{\prime}-\frac{3}{2}{\cal J}{\cal U}^{\prime}-\frac{1}{2}{\cal U}^{\prime\prime}.\label{eq:Miura2}
\end{align}
The Lagrange multipliers in the highest weight gauge, given by $\mu$
and $\nu$, are related to the variables in the diagonal gauge through
the following equations

\begin{align}
\zeta & =\mathcal{J}\mu-\mu^{\prime}-2\left(\mathcal{J}\mathcal{U}+\frac{1}{2}\mathcal{U}^{\prime}\right)\nu+\frac{1}{2}\mathcal{U}\nu^{\prime},\nonumber \\
\zeta_{\mathcal{U}} & =\mathcal{U}\mu-\left(\mathcal{J}^{2}-\mathcal{U}^{2}-\mathcal{J}^{\prime}\right)\nu+\frac{3}{2}\mathcal{J}\nu^{\prime}-\frac{1}{2}\nu^{\prime\prime}.\label{eq:zetas}
\end{align}
The complete Boussinesq hierarchy is then obtained from the mBoussinesq
one by virtue of the relations \eqref{eq:Miura2} and \eqref{eq:zetas}.
Indeed, from \eqref{eq:zetas}, one can prove that the chemical potentials
in the highest weight gauge take the form

\[
\mu=\frac{4\pi}{\kappa}\frac{\delta H_{\left(k\right)}^{\text{Bsq}}}{\delta\mathcal{L}},\quad\quad\nu=\frac{4\pi}{\kappa}\frac{\delta H_{\left(k\right)}^{\text{Bsq}}}{\delta\mathcal{W}},
\]
where $H_{\left(k\right)}^{\text{Bsq}}$ corresponds the $k$-th Hamiltonian
of the Boussinesq hierarchy (see appendix \ref{sec:7 Boussinesq-and-modified}
for more details on the Boussinesq hierarchy). For example, for the
first member given by $k=1$, one has

\[
H_{\left(1\right)}^{\text{Bsq}}=\frac{\kappa}{4\pi}\int d\phi\left(\lambda_{1}\mathcal{L}+\lambda_{2}\mathcal{W}\right),
\]
and hence the chemical potentials in the highest weight gauge become

\[
\mu=\lambda_{1},\quad\quad\nu=\lambda_{2}.
\]
In the particular case with $\lambda_{1}=1$ and $\lambda_{2}=0$
(first flow), the equations of motion are given by two chiral movers,
and the asymptotic conditions in eqs. \eqref{eq:Auxiliary connection HWg},
\eqref{eq:AtHWG} reduce to the ones in refs. \cite{Campoleoni:2010zq,Henneaux:2010xg}
but written in terms of the composite fields $\mathcal{L}$ and $\mathcal{W}$
according to \eqref{eq:Miura}. On the other hand, for the second
flow with $\lambda_{1}=0$ and $\lambda_{2}=1$, the field equation
in the bulk become equivalent to the Boussinesq equation in \eqref{eq:Bsqeq},
in agreement with the result found in ref. \cite{Perez:2016vqo}.

\section{Black holes\label{sec:3 Black holes}}

The line element is no longer gauge invariant in higher spin gravity,
since it generically changes under the action of a higher spin gauge
transformation. Therefore, the spacetime geometry and the causal structure
cannot be directly used to define black holes. In refs. \cite{Gutperle:2011kf,Ammon:2011nk},
a new notion of higher spin black hole was introduced in the Euclidean
formulation of the theory, by requiring trivial holonomies for the
gauge connection around a thermal cycle $\mathcal{C}$, i.e.,

\begin{equation}
\mathcal{H}_{\mathcal{C}}=\mathcal{P}e^{\int_{\mathcal{C}}a^{\pm}}=1.\label{eq:holocond}
\end{equation}
Here we have restored the $\pm$ superscript to denote the plus/minus
copy of the gauge field. If we assume that the Euclidean time is in
the range $0\leq t_{E}<1$ then, for time-independent configurations
in \eqref{eq:Auxiliary Connections}, the regularity condition \eqref{eq:holocond}
imposes the following restrictions on $\zeta^{\pm}$ and $\zeta_{\mathcal{U}}^{\pm}$
\begin{equation}
\zeta^{\pm}=\pi\left(2n+m\right)\,,\qquad\qquad\zeta_{\mathcal{U}}^{\pm}=\sqrt{3}\pi m,\label{eq:chemreg}
\end{equation}
with $m$ and $n$ being integers. Static configurations that obey
\eqref{eq:chemreg} are regular Euclidean solutions and consequently
we call them ``black holes.''

According to ref. \cite{Grumiller:2016kcp}, the entropy takes the
form
\begin{equation}
S=\frac{\kappa}{4}\int d\phi\left(\left(2n+m\right)\left(\mathcal{J}^{+}+\mathcal{J}^{-}\right)+\sqrt{3}m\left(\mathcal{U}^{+}+\mathcal{U}^{-}\right)\right),\label{eq:Entropy}
\end{equation}
which, by virtue of \eqref{eq:chemical potentials}, \eqref{eq:chemreg}
and \eqref{eq: decomp}, obeys the following first law 
\[
\delta S=\sum_{I=1}^{2}\left(\lambda_{I}^{+}\delta H_{\left(k\right)}^{I+}+\lambda_{I}^{-}\delta H_{\left(k\right)}^{I-}\right),
\]
where $H_{\left(k\right)}^{I\pm}$ are the left/right $k$-th Hamiltonian
of the mBoussinesq hierarchy associated to the flows labelled by $I=1,2$.
Note that the constants $\lambda_{I}^{\pm}$ correspond to the chemical
potentials conjugate to the extensive quantities $H_{\left(k\right)}^{I\pm}$.

When the integers $n$ and $m$ acquire the values $n=1$, $m=0$,
we obtain a branch which is connected with the pure gravitational
sector and the BTZ black hole. In that case, the entropy acquires
the simple expression 
\begin{equation}
S=\frac{\kappa}{2}\int d\phi\left(\mathcal{J}^{+}+\mathcal{J}^{-}\right).\label{eq:Entropy-1}
\end{equation}

As was pointed out in \cite{Grumiller:2016kcp}, for constants $\mathcal{J}^{\pm}$
and $\mathcal{U}^{\pm}$, the entropy for this branch acquires the
expected form found in \cite{Henneaux:2013dra,Bunster:2014mua} for
a higher spin black hole, once it is written in terms of the charges
of the $W$-algebra.

In sum, black holes that fit within our boundary conditions in eqs.
\eqref{eq:Auxiliary Connections} and \eqref{eq:chemical potentials}
are identified with static solutions of the $k$-th element of the
mBoussinesq hierarchy. It is worth noting that it is of fundamental
importance to consider both flows to admit generic black hole configurations
without restricting the possible space of solutions. To illustrate
some of their properties, we will study the particular cases with
$k=0,1,2$ in the branch connected with the BTZ black hole. For simplicity
we consider only the plus copy.

\emph{Case with $k=0$}. The general solution of the field equations
\eqref{eq:Field equations} that obeys \eqref{eq:chemreg} is given
by two arbitrary functions of $\phi$, i.e., $\mathcal{J}=\mathcal{J}\left(\phi\right)$
and $\mathcal{U}=\mathcal{U}\left(\phi\right)$. This case corresponds
to the ``higher spin black flower'' described in ref. \cite{Grumiller:2016kcp},
carrying an infinite set of $\hat{u}\left(1\right)$ soft hairy charges.
The particular configuration with constant $\mathcal{J}$ and vanishing
$\mathcal{U}$ in the branch with $n=1$, $m=0$, corresponds to the
BTZ black hole embedded within this set of boundary conditions.

\emph{Case with $k=1$}. The choice of boundary conditions \eqref{eq:chemical potentials},
together with the regularity conditions, imply the following differential
equations

\begin{eqnarray}
\lambda_{1}\mathcal{J}-\lambda_{2}\left(\mathcal{U}^{\prime}+2\mathcal{J}\mathcal{U}\right) & = & 2\pi,\nonumber \\
\lambda_{1}\mathcal{U}+\lambda_{2}\left(\mathcal{J}^{\prime}-\mathcal{J}^{2}+\mathcal{U}^{2}\right) & = & 0.\label{eq:systeq}
\end{eqnarray}
These equations relate the chemical potentials $\lambda_{1}$ and
$\lambda_{2}$ with the fields $\mathcal{J}$ and $\mathcal{U}$.
In particular, for constants $\mathcal{J}$ and $\mathcal{U}$ we
obtain
\begin{equation}
\lambda_{1}=\frac{2\pi\left(\mathcal{J}^{2}-\mathcal{U}^{2}\right)}{\mathcal{J}\left(\mathcal{J}^{2}-3\mathcal{U}^{2}\right)}\,,\qquad\qquad\lambda_{2}=\frac{2\pi\mathcal{U}}{\mathcal{J}\left(\mathcal{J}^{2}-3\mathcal{U}^{2}\right)}.\label{eq:lambdak1}
\end{equation}
The regularity conditions for the BTZ solution in the pure gravity
sector is recovered when $\mathcal{U}=0$, as expected. On the other
hand, the configurations with $\mathcal{J}\left(\mathcal{J}^{2}-3\mathcal{U}^{2}\right)=0$
possess non-trivial holonomies along the thermal cycle. Hence, one
might identify them with extremal configurations, along the lines
of ref. \cite{Henneaux:2015ywa} (see also \cite{Banados:2015tft}).
Indeed, when $\mathcal{U}^{\pm}=0$, solutions with vanishing $\mathcal{J}^{+}$
or $\mathcal{J}^{-}$ correspond to extremal BTZ black holes.

\emph{Case with $k=2$}. For constants $\mathcal{J}$ and $\mathcal{U}$,
the regularity conditions \eqref{eq:chemreg} fix the chemical potentials
$\lambda_{I}$ according to

\begin{eqnarray}
\lambda_{1} & = & \frac{3\pi\mathcal{U}\left(15\mathcal{J}^{4}-10\mathcal{J}^{2}\mathcal{U}^{2}+7\mathcal{U}^{4}\right)}{2\mathcal{J}\left(9\mathcal{J}^{8}-72\mathcal{J}^{6}\mathcal{U}^{2}+210\mathcal{J}^{4}\mathcal{U}^{4}-224\mathcal{J}^{2}\mathcal{U}^{6}-3\mathcal{U}^{8}\right)},\nonumber \\
\lambda_{2} & = & \frac{3\pi\left(3\mathcal{J}^{4}+6\mathcal{J}^{2}\mathcal{U}^{2}-5\mathcal{U}^{4}\right)}{2\mathcal{J}\left(9\mathcal{J}^{8}-72\mathcal{J}^{6}\mathcal{U}^{2}+210\mathcal{J}^{4}\mathcal{U}^{4}-224\mathcal{J}^{2}\mathcal{U}^{6}-3\mathcal{U}^{8}\right)}.\label{eq:lambdak2}
\end{eqnarray}
When $\mathcal{U}=0$, the auxiliary connection \eqref{eq:Auxiliary Connections}
reduces to the one that describes the BTZ geometry \cite{Afshar:2016wfy}.
However, in contrast to the case with $k=1$, the chemical potential
$\lambda_{1}$ associated to the first flow now vanishes, and hence,
the information of the BTZ black hole is completely encoded in the
second flow.

Note that for the cases with $k=1$ and $k=2$ described above, it
is of fundamental importance to take into account both flows to have
black holes characterized by two independent constants $\mathcal{J}^{\pm}$
and $\mathcal{U}^{\pm}$ for each copy. If one of the chemical potentials
$\lambda_{1}^{\pm}$ or $\lambda_{2}^{\pm}$ is set to zero, then
eqs. \eqref{eq:lambdak1} and \eqref{eq:lambdak2} would imply non-trivial
restrictions in the values of $\mathcal{J}^{\pm}$ and $\mathcal{U}^{\pm}$,
truncating in this way the spectrum of allowed black hole configurations.

\section{Hamiltonian reduction and boundary dynamics\label{sec:3 Reduction-of-the-Chern-Simons}}

In this section we will discuss the Hamiltonian reduction and the
boundary dynamics of the Chern-Simons action that describes spin-3
gravity endowed with the boundary conditions associated to the mBoussinesq
hierarchy. In the case of pure gravity with Brown-Henneaux asymptotic
conditions, the boundary dynamics is described by two left and right
chiral bosons which, by virtue of a Bäcklund transformation, turn
out to be equivalent to a Liouville theory \cite{Coussaert:1995zp,Henneaux:1999ib}.
The analysis was done by performing a Hamiltonian reduction of the
Wess-Zumino-Witten (WZW) theory at the boundary \cite{Forgacs:1989ac,Alekseev:1988ce,Witten:1989hf,Elitzur:1989nr}.

Here we follow an approach similar to the one proposed in ref. \cite{Gonzalez:2018jgp}
for KdV-type boundary conditions in pure gravity, where instead of
passing through the WZW theory, the boundary conditions are implemented
directly in the Hamiltonian action (see also \cite{Grumiller:2019tyl} for a 
different Hamiltonian reduction in the context of the KdV hierarchy).

Let us consider the Hamiltonian action with the appropriate boundary
term in eq. \eqref{eq:Action in hamiltonian form}. The constraints
$F_{ij}=0$ are locally solved by expressing the spatial components
of the gauge connection in terms of a group element $G$ as $A_{i}=G^{-1}\partial_{i}G$,
provided that there are no holes in the spatial section. After replacing
the solution of the constraints into the action \eqref{eq:Action in hamiltonian form},
the following decomposition is obtained

\begin{equation}
I_{can}\left[A\right]=I_{1}+I_{2}+B_{\infty},\label{eq: Total actions WZW}
\end{equation}
where

\begin{align}
I_{1} & =\frac{\kappa}{16\pi}\int dtdrd\phi\epsilon^{ij}\left\langle \partial_{t}\left(G^{-1}\right)\partial_{i}GG^{-1}\partial_{j}G\right\rangle ,\label{eq: I1 WZW action}\\
I_{2} & =-\frac{\kappa}{16\pi}\int d\phi dt\left\langle \partial_{t}G\partial_{\phi}\left(G^{-1}\right)\right\rangle .\label{eq:I2 WZW action}
\end{align}
The Wess-Zumino term $I_{1}$ reduces to a boundary term that vanishes
provided the group element is decomposed as $G=g\left(t,\phi\right)b\left(r\right)$
near the boundary (see appendix \ref{sec:Wess-Zumino-term}\textcolor{red}{{}
}for a detailed proof). Here $b\left(r\right)$ is the group element
that depends on the radial coordinate in eq. \eqref{eq:A Connections},
while $g\left(t,\phi\right)$ is such that the auxiliary connection
$a$ in \eqref{eq:Auxiliary Connections} is written as $a_{i}=g^{-1}\partial_{i}g$.
With this decomposition the term $I_{2}$ becomes

\begin{equation}
I_{2}=-\frac{\kappa}{16\pi}\int d\phi dt\left\langle \dot{g}\partial_{\phi}g^{-1}\right\rangle .\label{eq:I1 WZW 2}
\end{equation}
Since the auxiliary connection in \eqref{eq:Auxiliary Connections}
is a diagonal matrix, we can write the group element $g$ as follows

\begin{equation}
g=\exp\left[\sqrt{\frac{8\pi}{\kappa}}\varphi L_{0}+\sqrt{\frac{6\pi}{\kappa}}\psi W_{0}\right],\label{eq:Gauss decomposition}
\end{equation}
where $\varphi$ and $\psi$ are functions that depend only on $t$
and $\phi$. In what follows we will assume that these fields are
periodic in the angle, and consequently possible contributions coming
from non-trivial holonomies around $\phi$ are not considered in this
analysis.

Consistency with the asymptotic form of the auxiliary connection \eqref{eq:Auxiliary Connections}
then implies

\begin{equation}
\mathcal{J}=\sqrt{\frac{8\pi}{\kappa}}\varphi^{\prime}\,,\qquad\qquad\mathcal{U}=\sqrt{\frac{8\pi}{\kappa}}\psi^{\prime}.\label{eq: Dynamical fields of Gauss decomposition-2}
\end{equation}
Replacing \eqref{eq:Gauss decomposition} in \eqref{eq:I1 WZW 2}
we find

\begin{eqnarray}
I_{2} & = & \int d\phi dt\left(\varphi^{\prime}\dot{\varphi}+\psi^{\prime}\dot{\psi}\right).\label{eq:I1 WZW 3}
\end{eqnarray}
Thus, if we use the expression \eqref{eq:Boundary terms in terms of H}
for the boundary term $B_{\infty}$, we finally obtain the following
reduced action at the boundary

\begin{eqnarray}
I_{\left(k\right)} & = & \int dt\left[\int d\phi\left(\varphi^{\prime}\dot{\varphi}+\psi^{\prime}\dot{\psi}\right)-H_{\left(k\right)}\right].\label{eq:Total actions WZW 2-1-2-1}
\end{eqnarray}
This action describes the dynamics of the fields $\varphi$ and $\psi$,
whose interactions are described by the $k$-th Hamiltonian of the
mBoussinesq hierarchy. The members of the mBoussinesq hierarchy are
then recovered from the equations of motion derived from the action
\eqref{eq:Total actions WZW 2-1-2-1}, provided we identify the fields
according to \eqref{eq: Dynamical fields of Gauss decomposition-2}.
In this sense, the field equations coming from \eqref{eq:Total actions WZW 2-1-2-1}
define ``potential equations'' for the ones of the mBoussinesq hierarchy.

The action \eqref{eq:Total actions WZW 2-1-2-1} is invariant under
the following transformations
\[
\delta\varphi=\sqrt{\frac{\kappa}{8\pi}}\eta+f\left(t\right)\,,\qquad\qquad\delta\psi=\sqrt{\frac{\kappa}{8\pi}}\eta_{{\cal U}}+f_{{\cal U}}\left(t\right),
\]
where the parameter $\eta$ and $\eta_{{\cal U}}$, given by \eqref{eq:etas},
are associated to the infinite charges in involution of the integrable
system. Indeed, all the Hamiltonians of the hierarchy may be obtained
by a direct application of Noether theorem. On the other hand, the
arbitrary functions of the time $f\left(t\right)$ and $f_{{\cal U}}\left(t\right)$
define gauge symmetries of \eqref{eq:Total actions WZW 2-1-2-1} that
allow to gauge away the zero modes of these fields.

Furthermore, the action \eqref{eq:Total actions WZW 2-1-2-1} has
an additional Lifschitz scaling symmetry in the particular case when
only one of the two flows is considered. For the first flow, with
$\lambda_{1}=1$ and $\lambda_{2}=0$, the dynamical exponent is $z=3k-2$,
while for the second flow, with $\lambda_{1}=0$ and $\lambda_{2}=1$,
is $z=3k-1$, as expected from the invariance properties of the mBoussinesq
hierarchy described in section \ref{sec:Review-of-the}. For the $I$-th
flow of the $k$-th element of the hierarchy, the generators of the
Lifshitz algebra are given by

\[
H=H_{\left(k\right)}^{I}\,,\qquad\qquad P=H_{\left(1\right)}^{1}\,,\qquad\qquad D=-\frac{\kappa}{4\pi}\int d\phi\left(\frac{1}{2}\phi\left(\mathcal{J}^{2}+\mathcal{U}^{2}\right)\right)-ztH_{\left(k\right)}^{I}.
\]
Here, $H$ is the generators of translations in time, $P$ of translations
in space, and $D$ of anisotropic dilatations. Using the Dirac
brackets \eqref{eq:Dirac} it is straightforward to show that they
close in the Lifshitz algebra
\[
\left\{ P,H\right\} ^{\star}=0,\qquad\left\{ D,P\right\} ^{\star}=P,\qquad\left\{ D,H\right\} ^{\star}=zH,
\]
where $z$ is the dynamical exponent associated to the corresponding
flow.

Let us consider as an example the case with $k=1$. The action then
takes the following form

\begin{eqnarray*}
I_{\left(1\right)} & = & \int d\phi dt\left[\varphi^{\prime}\dot{\varphi}+\psi^{\prime}\dot{\psi}-\lambda_{1}\left(\varphi^{\prime2}+\psi^{\prime2}\right)+2\lambda_{2}\left(\varphi^{\prime}\psi^{\prime\prime}+\sqrt{\frac{8\pi}{\kappa}}\left(\varphi^{\prime2}\psi^{\prime}-\frac{1}{3}\psi^{\prime3}\right)\right)\right].
\end{eqnarray*}
For the flow with $\lambda_{1}=1$ and $\lambda_{2}=0$ we recover
the Floreanini-Jackiw action for two free chiral bosons \cite{Floreanini:1987as}.
On the other hand, for the flow with $\lambda_{1}=0$ and $\lambda_{2}=1$,
the action contains non-trivial interacting term and is invariant
under Lifshitz transformations with dynamical exponent $z=2$, as
expected.

\section{Some extensions: Generalized Gibbs ensemble, spin-N gravity and modified
Gelfand-Dickey hierarchies\label{sec:Some-extensions:-Generalized}}

\subsection{Generalized Gibbs Ensemble}

The existence of an infinite number of commuting charges in the mBoussinesq
hierarchy opens the possibility of study more general thermodynamic
ensembles that generically might include all possible charges. They
are called ``Generalized Gibbs Ensemble'' (GGE).

In the case of a two-dimensional conformal field theory, a GGE is
constructed with the infinite set of charges in involution of the
KdV hierarchy, which are obtained as composite operators in terms
of the stress tensor \cite{Sasaki:1987mm,Eguchi:1989hs,Bazhanov:1994ft}
(see refs. \cite{Calabrese:2011vdk,Sotiriadis:2014uza,PhysRevLett.115.157201,Vidmar_2016,deBoer:2016bov,Perez:2016vqo,Pozsgay_2017,Dymarsky:2018lhf,Maloney:2018hdg,Maloney:2018yrz,Dymarsky:2018iwx,Brehm:2019fyy,Dymarsky:2019etq,Dymarsky:2020tjh}
for recent results on GGE).

In our context, the Hamiltonians of the mBoussinesq hierarchy can
be used to construct the GGE of a two-dimensional CFT with spin-3
currents. As discussed in subsection \ref{subsec:Relation-with-the},
the Miura transformation \eqref{eq:Miura} maps the Hamiltonians of
the mBoussinesq hierarchy into the Hamiltonians of the Boussinesq
one, which depend on the stress tensor $\mathcal{L}$ and the spin-3
current $\mathcal{W}$. These Hamiltonians belong to the enveloping
algebra of the $W_{3}$-algebra, and consequently define an infinite
set of commuting charges that are composite operators in terms of
$\mathcal{L}$ and $\mathcal{W}$. In this sense, the relation between
the mBoussinesq hierarchy and spin-3 gravity discussed in this article,
provides a natural holographic bulk dual description of this type
of GGE.

This may be implemented as follows. Instead of considering one particular
$H_{\left(k\right)}$ as the Hamiltonian of the dynamical system,
we deal with a linear combination of them, i.e.,
\begin{equation}
H_{GGE}=\sum_{n=1}^{\infty}\sum_{I=1}^{2}\gamma_{n}\left(\lambda_{I}H_{\left(n\right)}^{I}\right).\label{eq:HGGE}
\end{equation}
If we want to interpret this general Hamiltonian as the one of a CFT$_{2}$
given by $H_{\left(1\right)}^{1}$, deformed by (multitrace) deformations
that include spin-3 currents, then we must identify the inverse (right)
temperature as $\beta_{+}=T_{+}^{-1}=\alpha_{1}\lambda_{1}$, and
the additional chemical potentials as $\mu_{n,I}:=T_{+}\alpha_{n}\lambda_{I}$.
However, this is not the only possibility. Any $H_{\left(k\right)}^{I}$
could be considered as the ``undeformed Hamiltonian,'' allowing
new branches that generically change the phase structure of the theory
\cite{Perezinprog}.

Black holes are described by static configurations of the dynamical
system with Hamiltonian \eqref{eq:HGGE}. The regularity condition
\eqref{eq:chemreg} then takes the form 
\[
\sum_{n=1}^{\infty}\sum_{I=1}^{2}\gamma_{n}\lambda_{I}\frac{\delta H_{\left(n\right)}^{I}}{\delta\mathcal{J}}=\frac{\kappa}{2}\,,\qquad\qquad\sum_{n=1}^{\infty}\sum_{I=1}^{2}\gamma_{n}\lambda_{I}\frac{\delta H_{\left(n\right)}^{I}}{\delta\mathcal{\mathcal{U}}}=0.
\]
These equations guarantee that the Euclidean action principle attains
an extremum, and hence they fully characterize the thermodynamics
of the GGE.

On the other hand, the boundary dynamics obtained by the Hamiltonian
reduction is easily generalized to the case when the Hamiltonian is
given by \eqref{eq:HGGE}. Indeed, the boundary action now becomes
\begin{eqnarray}
I_{GGE} & = & \int dt\left[\int d\phi\left(\varphi^{\prime}\dot{\varphi}+\psi^{\prime}\dot{\psi}\right)-H_{GGE}\right].\label{eq:Total actionGGE}
\end{eqnarray}

\subsection{Higher spin gravity with gauge group $SL\left(N,\mathbb{R}\right)\times SL\left(N,\mathbb{R}\right)$
and modified Gelfand-Dickey hierarchies}

Our results can be generalized to the case of three-dimensional higher
spin gravity with gauge group $SL\left(N,\mathbb{R}\right)\times SL\left(N,\mathbb{R}\right)$,
where the corresponding integrable systems are the called ``modified
Generalized KdV hierarchies,'' or ``modified Gelfand-Dickey (mGD)
hierarchies'' (see e.g. chapter 4 of ref. \cite{2003ASMP...26.....D}).
The link with higher spin gravity is based on the zero curvature formulation
of these integrable systems \cite{Drinfeld:1984qv}.

Let us consider asymptotic conditions for spin-$N$ gravity described
by the following auxiliary gauge connection valued on the $sl\left(N,\mathbb{R}\right)$
algebra
\begin{equation}
a=\left({\cal J}d\phi+\zeta dt\right)L_{0}+\sum_{s=3}^{N}\sigma_{s}\left(\mathcal{U}^{\left(s\right)}d\phi+\zeta_{\mathcal{U}}^{\left(s\right)}dt\right)W_{0},\label{eq:Auxiliary Connections SLN}
\end{equation}
with $\sigma_{s}$ given by 
\[
\sigma_{s}=\left(\frac{\left(2s-1\right)!\left(2s-2\right)!}{48(s-1)!^{4}}\frac{1}{\prod_{i=2}^{s-1}\left(N^{2}-i^{2}\right)}\right)^{\frac{1}{2}}.
\]
The Chern-Simons action takes the same form as in \eqref{eq:Chern-Simons Action}, 
where one has to replace $\kappa_{3}\rightarrow\kappa_{N}=3l/(2N(N^{2}-1)G)$.
Hence, the variation of the boundary term of the canonical Chern-Simons
action becomes
\begin{equation}
\delta B_{\infty}=-\frac{\kappa}{4\pi}\int dtd\phi\left(\zeta\delta\mathcal{J}+\sum_{s=3}^{N}\zeta_{\mathcal{U}}^{\left(s\right)}\delta\mathcal{U}^{\left(s\right)}\right).\label{eq:Variation of boundary terms-1}
\end{equation}
To make contact with the mGD hierarchies, we choose the Lagrange multipliers
as follows
\begin{equation}
\zeta=\frac{4\pi}{\kappa}\frac{\delta H_{\left(k,N\right)}^{\text{mGD}}}{\delta\mathcal{J}},\qquad\qquad\zeta_{\mathcal{U}}^{\left(s\right)}=\frac{4\pi}{\kappa}\frac{\delta H_{\left(k,N\right)}^{\text{mGD}}}{\delta\mathcal{U}^{\left(s\right)}},\label{eq:chemical potentials-1}
\end{equation}
where $H_{\left(k,N\right)}^{\text{mGD}}$ corresponds to the $k$-th
Hamiltonian of the $N$-th hierarchy \cite{Drinfeld:1984qv}. With
this choice of boundary conditions, the boundary term of the canonical
Chern-Simons action integrates as 
\[
B_{\infty}=-\int dtH_{\left(k,N\right)}.
\]
As expected, the Hamiltonian of the higher spin theory coincides with
one of the hierarchy.

The Dirac brackets are described by $N-1$ $\hat{u}\left(1\right)$
current algebras
\begin{eqnarray}
\left\{ \mathcal{J}\left(\phi\right),\mathcal{J}\left(\phi^{\prime}\right)\right\} ^{\star} & = & \frac{4\pi}{\kappa}\partial_{\phi}\delta\left(\phi-\phi^{\prime}\right),\nonumber \\
\left\{ \mathcal{U}^{\left(s\right)}\left(\phi\right),\mathcal{U}^{\left(s'\right)}\left(\phi^{\prime}\right)\right\} ^{\star} & = & \frac{4\pi}{\kappa}\partial_{\phi}\delta\left(\phi-\phi^{\prime}\right)\delta^{s,s'},\label{eq:Dirac-1}
\end{eqnarray}
matching the first Poisson structure of the mGD hierarchies. Moreover,
their members are obtained from the equations of motion of the higher
spin theory with the boundary conditions \eqref{eq:Auxiliary Connections SLN},
\eqref{eq:chemical potentials-1}, and are given by
\[
\dot{\mathcal{J}}=\frac{4\pi}{\kappa}\partial_{\phi}\left(\frac{\delta H_{\left(k,N\right)}^{\text{mGD}}}{\delta\mathcal{J}}\right)\,,\qquad\qquad\dot{\mathcal{U}}^{\left(s\right)}=\frac{4\pi}{\kappa}\partial_{\phi}\left(\frac{\delta H_{\left(k,N\right)}^{\text{mGD}}}{\delta\mathcal{U}^{\left(s\right)}}\right).
\]

The mGD hierarchies are related to the called ``Generalized KdV hierarchies,''
or ``Gelfand-Dickey (GD) hierarchies'' by an appropriate generalization
of the Miura transformation. One of the two Poisson brackets of the
GD hierarchies is described by the $W_{N}$-algebra, whose generators
are composite in terms of the $\hat{u}\left(1\right)$ currents of
the mGD hierarchies. Hence, according to the Hamiltonian reduction
in \cite{Drinfeld:1984qv}, the generalized Miura transformation should
emerge geometrically from our boundary conditions once they are expressed
in the highest weight gauge, as in the case for $N=3$ described in
subsection \ref{subsec:Relation-with-the}. Note that since generically
the expression for the stress tensor of the $W_{N}$-algebra in terms
of $\hat{u}\left(1\right)$ currents is the one of a twisted Sugawara
construction, there is a particular flow in which the currents are
chiral. This case precisely corresponds to one of the proposals in
ref. \cite{Perez:2020klz}, for describing gravitational duals of
averaged CFT's on the Narain lattice \cite{Maloney:2020nni,Afkhami-Jeddi:2020ezh}
(see \cite{Cotler:2020ugk} for an alternative proposal for a possible
gravitational dual).

\section{Final remarks\label{sec:5 Final-remarks}}

We have shown that the dynamics of three-dimensional higher spin gravity
with gauge group $SL\left(N,\mathbb{R}\right)\times SL\left(N,\mathbb{R}\right)$
endowed with a certain special class of boundary conditions in the
diagonal gauge, reduces to the one of the integrable system corresponding
to the $N$-th modified Gelfand-Dickey hierarchy. The particular case
with $N=3$ is associated to the mBoussinesq hierarchy, whose first
member defines a potential equation for the Boussinesq one, which
was found for the first time in the nineteenth century in the context
of the study of solitary waves (solitons) in fluid dynamics. In this
sense, it would be interesting to explore the possibility that these
results could be understood in the context of the fluid-gravity correspondence
\cite{Bhattacharyya:2008jc,Haack:2008cp,Bhattacharyya:2008mz,Hubeny:2011hd},
along the lines of \cite{Campoleoni:2018ltl}.

On the other hand, in section \ref{sec:3 Reduction-of-the-Chern-Simons}
we studied the Hamiltonian reduction and the boundary dynamics induced
by our choice of boundary conditions. In this analysis it was explicitly
assumed that the fields $\varphi$ and $\psi$ were periodic in the
angle $\phi$, leading to a dynamics described by the action \eqref{eq:Total actions WZW 2-1-2-1}.
However, this assumption does not take into account the black hole
configurations described in section \ref{sec:3 Black holes}. This
suggests the possibility of trying to introduce non-trivial holonomies
around the non-contractible cycle along the lines of refs. \cite{Henneaux:2019sjx},
to incorporate these solutions into the analysis.

The entire integrable structure of the family of (m)GD hierarchies
can be embedded in a single universal (2+1)-dimensional integrable
system called the Kadomtsev-Petviashvili (KP) hierarchy \cite{1970SPhD...15..539K}.
Many different (1+1)-dimensional integrable systems, including the
(m)GD hierarchies, are recovered for some specific (dimensional) reductions
of the KP hierarchy (see e.g. \cite{2003ASMP...26.....D}). It would
then be natural to explore the possibility that the KP hierarchy could
emerge from the asymptotic structure of a higher dimensional bulk
(higher spin) gravitational theory.

Other extensions of our results might include the possible relation
of certain two-dimensional integrable systems with three-dimensional
higher spin gravity with vanishing cosmological constant \cite{Afshar:2013vka,Gonzalez:2013oaa,Gary:2014ppa,Matulich:2014hea,Ammon:2017vwt},
generalizations in the context of hypergravity \cite{Henneaux:2015ywa,Fuentealba:2015jma,Fuentealba:2015wza,Henneaux:2015tar}, or extensions to  generalized Boussinesq hierarchies as the ones described in ref. \cite{Antonowicz_1991}.

\acknowledgments{We thank F\'abio Novaes and Ricardo Troncoso for some useful discussions. The work of EO was  funded by the PhD grant CONICYT-PCHA/Doctorado Nacional/2016-21161352. This research has been partially supported by Fondecyt grants N$\textsuperscript{\underline{o}}$ 1171162, 1181496, 1181031. The Centro de Estudios Cient\'ificos (CECs) is funded by the Chilean Government through the Centers of Excellence Base Financing Program of Conicyt.}

\appendix

\section{Second Hamiltonian structure of the modified Boussinesq hierarchy\label{sec:Second-Hamiltonian-structure}}

The second Hamiltonian structure of the mBoussinesq hierarchy is defined
by the operator $\mathcal{D}_{\left(2\right)}$ in \eqref{eq:secondPstr},
whose explicit components are given by

\begin{eqnarray*}
\frac{\hat{\kappa}}{4\pi}\mathcal{D}_{(2)}^{11} & = & -4\left(2\left({\cal U}^{\prime\prime}+2\left({\cal J}{\cal U}\right)^{\prime}\right)\partial_{\phi}^{-1}\left({\cal J}\partial_{\phi}+\partial_{\phi}^{2}\right)+{\cal J}^{\prime}\partial_{\phi}^{-1}\left(\left(4{\cal J}\mathcal{U}+3\mathcal{U}^{\prime}\right)\partial_{\phi}+\mathcal{U}\partial_{\phi}^{2}\right)\right.\\
 &  & \left.+\left(8{\cal J}^{2}\mathcal{U}-5{\cal U}\mathcal{J}^{\prime}-3{\cal U}^{\prime\prime}\right)\partial_{\phi}-3\mathcal{U}^{\prime}\partial_{\phi}^{2}-2{\cal U}\partial_{\phi}^{3}\right),\\
\frac{\hat{\kappa}}{4\pi}\mathcal{D}_{(2)}^{12} & = & -4\left(2\left({\cal U}^{\prime\prime}+2\left({\cal J}{\cal U}\right)^{\prime}\right)\partial_{\phi}^{-1}\left({\cal U}\partial_{\phi}\right)+{\cal J}^{\prime}\partial_{\phi}^{-1}\left(\left(2\mathcal{J}^{2}-2\mathcal{U}^{2}+\mathcal{J}^{\prime}\right)\partial_{\phi}+3\mathcal{J}\partial_{\phi}^{2}+\partial_{\phi}^{3}\right)\right.\\
 &  & \left.+\left(2{\cal J}^{3}+2{\cal J}\mathcal{U}^{2}+2{\cal U}^{2\prime}-3{\cal J}{\cal J}^{\prime}-{\cal J}^{\prime\prime}\right)\partial_{\phi}+\left(\mathcal{J}^{2}+\mathcal{U}^{2}-4\mathcal{J}^{\prime}\right)\partial_{\phi}^{2}-2{\cal J}\partial_{\phi}^{3}-\partial_{\phi}^{4}\right),\\
\frac{\hat{\kappa}}{4\pi}\mathcal{D}_{(2)}^{21} & = & -4\left(2\left({\cal J}^{2\prime}-{\cal U}^{2\prime}-{\cal J}^{\prime\prime}\right)\partial_{\phi}^{-1}\left({\cal J}\partial_{\phi}+\partial_{\phi}^{2}\right)+{\cal U}^{\prime}\partial_{\phi}^{-1}\left(\left(4{\cal J}\mathcal{U}+3\mathcal{U}^{\prime}\right)\partial_{\phi}+\mathcal{U}\partial_{\phi}^{2}\right)\right.\\
 &  & \left.+\left(2{\cal J}^{3}+2{\cal J}\mathcal{U}^{2}-4{\cal J}^{2\prime}+3{\cal U}{\cal U}^{\prime}+{\cal J}^{\prime\prime}\right)\partial_{\phi}-\left(3\mathcal{J}^{\prime}+\mathcal{J}^{2}+\mathcal{U}^{2}\right)\partial_{\phi}^{2}-2{\cal J}\partial_{\phi}^{3}+\partial_{\phi}^{4}\right),\\
\frac{\hat{\kappa}}{4\pi}\mathcal{D}_{(2)}^{22} & = & -4\left(-2\left({\cal J}^{\prime\prime}+{\cal U}^{2\prime}-{\cal J}^{2\prime}\right)\partial_{\phi}^{-1}\left({\cal U}\partial_{\phi}\right)+{\cal U}^{\prime}\partial_{\phi}^{-1}\left(\left(2\mathcal{J}^{2}-2\mathcal{U}^{2}+\mathcal{J}^{\prime}\right)\partial_{\phi}+3\mathcal{J}\partial_{\phi}^{2}+\partial_{\phi}^{3}\right)\right.\\
 &  & \left.+\left(-4\mathcal{U}^{3}+4{\cal J}^{2}\mathcal{U}-3{\cal J}{\cal U}^{\prime}-4{\cal J}^{\prime}{\cal U}+{\cal U}^{\prime\prime}\right)\partial_{\phi}+2\mathcal{U}^{\prime}\partial_{\phi}^{2}+2{\cal U}\partial_{\phi}^{3}\right).
\end{eqnarray*}

\section{Gelfand-Dickey polynomials and Hamiltonians of the modified Boussinesq
hierarchy\label{sec:8 Gelfand-Dikii-polynomials-and}}

In this appendix we exhibit the first Gelfand-Dickey polynomials and
Hamiltonians of the mBoussinesq hierarchy.

The Gelfand-Dickey polynomials can be explicitly constructed using
the recurrence relation \eqref{eq:rec}. The first of them are given
by

\begin{eqnarray*}
\frac{4\pi}{\hat{\kappa}}R_{\mathcal{J}}^{\left(0\right)} & = & \lambda_{1},\\
\frac{4\pi}{\hat{\kappa}}R_{\mathcal{J}}^{\left(1\right)} & = & \lambda_{1}\mathcal{J}-\lambda_{2}\left(\mathcal{U}^{\prime}+2\mathcal{J}\mathcal{U}\right),\\
\frac{4\pi}{\hat{\kappa}}R_{\mathcal{J}}^{\left(2\right)} & = & 4\lambda_{1}\left(-4\mathcal{J}^{3}\mathcal{U}-\frac{4}{3}\mathcal{J}\mathcal{U}^{3}-2\mathcal{J}^{2}\mathcal{U}^{\prime}-2\mathcal{U}^{2}\mathcal{U}^{\prime}+2\mathcal{J}^{\prime}\mathcal{U}^{\prime}+2\mathcal{J}^{\prime\prime}\mathcal{U}+2\mathcal{J}\mathcal{U}^{\prime\prime}\right.\\
{\color{red}} & {\color{red}} & \left.+\mathcal{U}^{\prime\prime\prime}\right)+4\lambda_{2}\left(\mathcal{J}^{5}+10\mathcal{J}^{3}\mathcal{U}^{2}-\frac{5}{3}\mathcal{J}\mathcal{U}^{4}-5\mathcal{J}\mathcal{J}^{\prime2}+5\mathcal{J}^{2}\mathcal{U}^{2\prime}-\frac{10}{12}\mathcal{U}^{4\prime}\right.\\
 &  & \left.-5\mathcal{J}^{\prime}\mathcal{U}^{2\prime}+5\mathcal{J}\mathcal{U}^{\prime2}-5\mathcal{J}^{2}\mathcal{J}^{\prime\prime}-5\mathcal{J}^{\prime\prime}\mathcal{U}^{2}-5\mathcal{J}^{\prime}\mathcal{J}^{\prime\prime}+5\mathcal{U}^{\prime}\mathcal{U}^{\prime\prime}+\mathcal{J}^{\prime\prime\prime\prime}\right),
\end{eqnarray*}
\begin{eqnarray*}
\frac{4\pi}{\hat{\kappa}}R_{\mathcal{U}}^{\left(0\right)} & = & \lambda_{2},\\
\frac{4\pi}{\hat{\kappa}}R_{\mathcal{U}}^{\left(1\right)} & = & \lambda_{1}\mathcal{U}+\lambda_{2}\left(\mathcal{J}^{\prime}-\mathcal{J}^{2}+\mathcal{U}^{2}\right),\\
\frac{4\pi}{\hat{\kappa}}R_{\mathcal{U}}^{\left(2\right)} & = & 4\lambda_{1}\left(-\mathcal{J}^{4}-2\mathcal{J}^{2}\mathcal{U}^{2}+\frac{5}{3}\mathcal{U}^{4}+\frac{2}{3}\mathcal{J}^{3\prime}+2\mathcal{U}^{2}\mathcal{J}^{\prime}+\mathcal{J}^{\prime2}-\mathcal{U}^{\prime2}+2\mathcal{J}\mathcal{J}^{\prime\prime}\right.\\
{\color{red}} & {\color{red}} & \left.-2\mathcal{U}\mathcal{U}^{\prime\prime}-\mathcal{J}^{\prime\prime\prime}\right)+4\lambda_{2}\left(5\mathcal{J}^{4}\mathcal{U}-\frac{10}{3}\mathcal{J}^{2}\mathcal{U}^{3}+\frac{7}{3}\mathcal{U}^{5}-\frac{10}{3}\mathcal{J}^{3\prime}\mathcal{U}+\frac{10}{3}\mathcal{J}^{\prime}\mathcal{U}^{3}\right.\\
{\color{red}} & {\color{red}} & \left.+5\mathcal{U}\mathcal{J}^{\prime2}-5\mathcal{J}^{2\prime}\mathcal{U}^{\prime}-5\mathcal{U}\mathcal{U}^{\prime2}+5\mathcal{J}^{\prime\prime}\mathcal{U}^{\prime}-5\mathcal{J}^{2}\mathcal{U}^{\prime\prime}-5\mathcal{U}^{2}\mathcal{U}^{\prime\prime}+5\mathcal{J}^{\prime}\mathcal{U}^{\prime\prime}+\mathcal{U}^{\prime\prime\prime\prime}\right).
\end{eqnarray*}
The corresponding Hamiltonians can then be obtained using eq. \eqref{eq:GelfandDikii}.
Thus,

\begin{eqnarray*}
\frac{4\pi}{\hat{\kappa}}H_{\left(0\right)} & = & \int d\phi\left(\lambda_{1}\mathcal{J}+\lambda_{2}\mathcal{U}\right),\\
\frac{4\pi}{\hat{\kappa}}H_{\left(1\right)} & = & \int d\phi\left\{ \frac{\lambda_{1}}{2}\left(\mathcal{J}^{2}+\mathcal{U}^{2}\right)+\lambda_{2}\left(\frac{1}{3}\mathcal{U}^{3}-\mathcal{J}^{2}\mathcal{U}-\mathcal{J}\mathcal{U}^{\prime}\right)\right\} ,\\
\frac{4\pi}{\hat{\kappa}}H_{\left(2\right)} & = & \int d\phi\left\{ \frac{4\lambda_{1}}{3}\left(\mathcal{U}^{5}-3\mathcal{J}^{4}\mathcal{U}-2\mathcal{J}^{2}\mathcal{U}^{3}-3\mathcal{J}^{\prime2}\mathcal{U}-2\mathcal{J}^{3}\mathcal{U}^{\prime}-2\mathcal{J}\mathcal{U}^{3\prime}\right.\right.\\
{\color{red}} & {\color{red}} & \left.+3\mathcal{J}^{2}\mathcal{U}^{\prime\prime}-\frac{3}{2}\mathcal{U}^{2}\mathcal{U}^{\prime\prime}+3\mathcal{J}\mathcal{U}^{\prime\prime\prime}\right)+\frac{2\lambda_{2}}{3}\left(\mathcal{J}^{6}+15\mathcal{J}^{4}\mathcal{U}^{2}-5\mathcal{J}^{2}\mathcal{U}^{4}+\frac{7}{3}\mathcal{U}^{6}\right.\\
 &  & \left.+15\mathcal{J}^{2}\mathcal{J}^{\prime2}+15\mathcal{U}^{2}\mathcal{J}^{\prime2}+10\mathcal{J}^{3}\mathcal{U}^{2\prime}-5\mathcal{J}\mathcal{U}^{4\prime}+15\mathcal{J}^{2}\mathcal{U}^{\prime2}+15\mathcal{U}^{2}\mathcal{U}^{\prime2}\right.\\
 &  & \left.\left.-10\mathcal{J}\mathcal{J}^{\prime}\mathcal{J}^{\prime\prime}+30\mathcal{J}\mathcal{U}^{\prime}\mathcal{U}^{\prime\prime}+3\mathcal{J}\mathcal{J}^{\prime\prime\prime\prime}+3\mathcal{U}\mathcal{U}^{\prime\prime\prime\prime}\right)\right\} .
\end{eqnarray*}

\section{Boussinesq hierarchy\label{sec:7 Boussinesq-and-modified}}

The Boussinesq hierarchy is an integrable bi-Hamiltonian system which
possesses two different Poisson brackets defined by the following
operators
\begin{equation}
\mathcal{D}_{\left(1\right)}^{\text{Bsq}}=\frac{\pi}{2\hat{\kappa}}\left(\begin{array}{cc}
0 & \partial_{\phi}\\
\partial_{\phi} & 0
\end{array}\right),\label{eq:Bq Operators}
\end{equation}
\[
\mathcal{D}_{\left(2\right)}^{\text{Bsq}}=\frac{4\pi}{\hat{\kappa}}\begin{pmatrix}2{\cal L}\partial_{\phi}+{\cal L}^{\prime}-\partial_{\phi}^{3} & 3\mathcal{W}\partial_{\phi}+2\mathcal{W}^{\prime}\\
3\mathcal{W}\partial_{\phi}+\mathcal{W}^{\prime} & -\frac{1}{2}\mathcal{L}^{\prime\prime\prime}+2\mathcal{L}^{2\prime}-\frac{9}{4}\left(\mathcal{L}^{\prime\prime}-\frac{16}{9}\mathcal{L}^{2}\right)\partial_{\phi}-\frac{15}{4}{\cal L}^{\prime}\partial_{\phi}^{2}-\frac{5}{2}{\cal L}\partial_{\phi}^{3}+\frac{1}{4}\partial_{\phi}^{5}
\end{pmatrix}.
\]
The Poisson bracket associated to the operator $\mathcal{D}_{\left(2\right)}^{\text{Bsq}}$
is given by the classical $W_{3}$-algebra.

The infinite Hamiltonians in involution can be obtained using the
following recursion relation
\[
\mathcal{D}_{(1)}^{\text{Bsq}}\left(\begin{array}{c}
R_{\mathcal{L}}\\
R_{\mathcal{W}}
\end{array}\right)_{\left(k+1\right)}=\mathcal{D}_{(2)}^{\text{Bsq}}\left(\begin{array}{c}
R_{\mathcal{L}}\\
R_{\mathcal{W}}
\end{array}\right)_{\left(k\right)}.
\]
Here the corresponding Gelfand-Dickey polynomials are defined through
\[
\left(\begin{array}{c}
R_{\mathcal{L}}\\
R_{\mathcal{W}}
\end{array}\right)_{\left(k\right)}=\begin{pmatrix}\frac{\delta H_{\left(k\right)}^{\text{Bsq}}}{\delta\mathcal{L}}\\
\frac{\delta H_{\left(k\right)}^{\text{Bsq}}}{\delta\mathcal{\mathcal{W}}}
\end{pmatrix},
\]
where the first Hamiltonian is given by

\[
H_{\left(1\right)}^{\text{Bsq}}=\frac{\hat{\kappa}}{4\pi}\int d\phi\left(\lambda_{1}\mathcal{L}+\lambda_{2}\mathcal{W}\right).
\]
Therefore, the members of the hierarchy can be written as follows

\begin{equation}
\left(\begin{array}{c}
\dot{\mathcal{L}}\\
\dot{\mathcal{W}}
\end{array}\right)_{\left(k\right)}=\mathcal{D}_{(1)}^{\text{Bsq}}\left(\begin{array}{c}
R_{\mathcal{L}}\\
R_{\mathcal{W}}
\end{array}\right)_{\left(k+1\right)}=\mathcal{D}_{(2)}^{\text{Bsq}}\left(\begin{array}{c}
R_{\mathcal{L}}\\
R_{\mathcal{W}}
\end{array}\right)_{\left(k\right)}.\label{eq:Bq Hierarchy}
\end{equation}

As explained in section \ref{sec:Review-of-the}, the Boussinesq and
the mBoussinesq hierarchies are related by the Miura transformation
\eqref{eq:Miura}, that can be rewritten in the following vector form

\[
\left(\begin{array}{c}
\mathcal{L}\\
\mathcal{W}
\end{array}\right)=F\left[{\cal J},{\cal U}\right],
\]
for a functional $F$ defined through \eqref{eq:Miura}. Taking the
derivative with respect to the time one obtains

\begin{equation}
\left(\begin{array}{c}
\dot{\mathcal{L}}\\
\dot{\mathcal{W}}
\end{array}\right)=M\left(\begin{array}{c}
\dot{\mathcal{J}}\\
\dot{{\cal U}}
\end{array}\right),\label{eq:Miura vector}
\end{equation}
where $M=M\left[{\cal J},{\cal U}\right]$ correspond to the Fr\'echet derivative of $F$ with respect to ${\cal J}$ and ${\cal U}$ \cite{Mathieu:1991},
and is precisely given by the matrix $M$ in \eqref{eq:M}, i.e.,

\[
M=\left(\begin{array}{cc}
{\cal J}+\partial_{\phi}\qquad & \mathcal{U}\\
-2{\cal J}\mathcal{U}-\frac{1}{2}\mathcal{U}\partial_{\phi}-\frac{3}{2}{\cal U}^{\prime}\qquad & \mathcal{U}^{2}-{\cal J}^{2}-\frac{1}{2}{\cal J}^{\prime}-\frac{3}{2}{\cal J}\partial_{\phi}-\frac{1}{2}\partial_{\phi}^{2}
\end{array}\right).
\]
If one takes into account its formal adjoint

\[
M^{\dagger}=\left(\begin{array}{cc}
\;\;{\cal J}-\partial_{\phi}\qquad & -2{\cal J}\mathcal{U}+\frac{1}{2}\mathcal{U}\partial_{\phi}-{\cal U}^{\prime}\\
\;\;\mathcal{U}\qquad & \mathcal{U}^{2}-{\cal J}^{2}+{\cal J}^{\prime}+\frac{3}{2}{\cal J}\partial_{\phi}-\frac{1}{2}\partial_{\phi}^{2}
\end{array}\right),
\]
the Gelfand-Dickey polynomials of both hierarchies are then related
by

\begin{equation}
\left(\begin{array}{c}
R_{\mathcal{J}}\\
R_{\mathcal{U}}
\end{array}\right)=M^{\dagger}\left(\begin{array}{c}
R_{\mathcal{L}}\\
R_{\mathcal{W}}
\end{array}\right).\label{eq:Miura Gelfand-Dikii}
\end{equation}
Taking into account eqs. \eqref{eq:Miura vector}, \eqref{eq:EOM}
and \eqref{eq:Miura Gelfand-Dikii} we can write

\[
\left(\begin{array}{c}
\dot{\mathcal{L}}\\
\dot{\mathcal{W}}
\end{array}\right)_{\left(k\right)}=M\mathcal{D}\left(\begin{array}{c}
R_{\mathcal{J}}\\
R_{\mathcal{U}}
\end{array}\right)_{\left(k\right)}=M\mathcal{D}M^{\dagger}\left(\begin{array}{c}
R_{\mathcal{L}}\\
R_{\mathcal{W}}
\end{array}\right)_{\left(k\right)}=\mathcal{D}_{(2)}^{\text{Bsq}}\left(\begin{array}{c}
R_{\mathcal{L}}\\
R_{\mathcal{W}}
\end{array}\right)_{\left(k\right)},
\]
which imply that the second Poisson structure for the Boussinesq hierarchy
can be expressed in terms of the first Poisson structure of the mBoussinesq
hierarchy according to
\[
\mathcal{D}_{(2)}^{\text{Bsq}}=M\mathcal{D}M^{\dagger}.
\]

\section{Fundamental representation of the principal embedding of $sl(2,\mathbb{R})$
within $sl(N,\mathbb{R})$\label{sec:6 SL(3,R)-in-the}}

In the principal embedding of the $sl(2,\mathbb{R})$ algebra within
the $sl(N,\mathbb{R})$ algebra the generators can be written in the
basis \emph{$\left\{ L_{i},W_{m}^{(s)}\right\} $}, with $i=-1,0,1$,
$s=3,4,\dots$ and $m=-s+1,\dots,s-1$. In the fundamental representation
of $sl(N,\mathbb{R})$, the generators may be represented by the following
$N\times N$ matrices

\[
\begin{aligned}\left(L_{1}\right)_{jk} & =-\sqrt{j\left(N-j\right)}\delta_{j+1,k},\\
\left(L_{-1}\right)_{jk} & =\sqrt{k\left(N-k\right)}\delta_{j,k+1},\\
\left(L_{0}\right)_{jk} & =\frac{1}{2}\left(N+1-2j\right)\delta_{j,k},
\end{aligned}
\]
\[
\begin{aligned}W_{m}^{(s)} & =2\left(-1\right)^{s-m-1}\frac{\left(s+m-1\right)!}{\left(2s-2\right)!}\underbrace{\left[L_{-1},\left[L_{-1},\cdots\left[L_{-1},\left(L_{1}\right)^{s-1}\right]\cdots\right]\right]}_{s-m-1\text{ terms }},\\
 & =2\left(-1\right)^{s-m-1}\frac{\left(s+m-1\right)!}{\left(2s-2\right)!}\left(\mathrm{ad}_{L_{-1}}\right)^{s-m-1}\left(L_{1}\right)^{s-1}.
\end{aligned}
\]
with $j,k=1,\dots,N$, and where \emph{$\mathrm{ad}_{\mathrm{x}}\left(\mathrm{Y}\right):=\left[\mathrm{X},\mathrm{Y}\right]$.}
From the commutation relations
\[
\begin{aligned}\left[L_{i},L_{j}\right] & =\left(i-j\right)L_{i+j},\\
\left[L_{i},W_{m}^{(s)}\right] & =\left(\left(s-1\right)i-m\right)W_{i+m}^{(s)},
\end{aligned}
\]
it can be seen that the $L_{i}$ generators close in a $sl(2,\mathbb{R})$
subalgebra, while the generators $W_{m}^{(s)}$ transform in a spin-$s$
representation under $sl(2,\mathbb{R})$.

\section{Wess-Zumino term\label{sec:Wess-Zumino-term}}

Here we show that for our boundary conditions in \eqref{eq:A Connections},
\eqref{eq:Auxiliary Connections}, \eqref{eq:chemical potentials},
the Wess-Zumino term 
\begin{align*}
I_{1} & =\frac{\kappa}{16\pi}\int dtdrd\phi\epsilon^{ij}\left\langle \partial_{t}\left(G^{-1}\right)\partial_{i}GG^{-1}\partial_{j}G\right\rangle ,
\end{align*}
 in \eqref{eq: I1 WZW action} vanishes.

Let us perform the following Gauss decomposition of the group element

\begin{equation}
G=e^{TL_{1}+MW_{1}+QW_{2}}e^{\Phi L_{0}+\Phi_{W}W_{0}}e^{XL_{-1}+YW_{-1}+ZW_{-2}},\label{eq:GWZW}
\end{equation}
where all the functions that appear in \eqref{eq:GWZW} generically
depend on $t$, $r$ and $\phi$. Then, if we replace \eqref{eq:GWZW}
in $I_{1}$, one can show that it reduces to a boundary term of the
form

\begin{eqnarray}
I_{1} & = & \frac{\kappa}{16\pi}\int d\phi dt\left[2e^{\Phi+2\Phi_{W}}\left(\left(X^{\prime}+Y^{\prime}\right)\left(\dot{M}+\dot{T}\right)-\left(\dot{X}+\dot{Y}\right)\left(M^{\prime}+T^{\prime}\right)\right)\right.\nonumber \\
 &  & \left.-2e^{\Phi-2\Phi_{W}}\left(\left(X^{\prime}-Y^{\prime}\right)\left(\dot{M}-\dot{T}\right)-\left(\dot{X}-\dot{Y}\right)\left(M^{\prime}-T^{\prime}\right)\right)\right.\label{eq:I2ap}\\
 &  & +e^{2\Phi}\left(8\left(XY^{\prime}-YX^{\prime}\right)\dot{Q}-8\left(X\dot{Y}-Y\dot{X}\right)Q^{\prime}+8\left(T\dot{M}-M\dot{T}\right)Z^{\prime}-8\left(TM^{\prime}-MT^{\prime}\right)\dot{Z}\right.\nonumber \\
 &  & \left.\left.+4\left(X\dot{Y}-Y\dot{X}\right)\left(TM^{\prime}-MT^{\prime}\right)-4\left(XY^{\prime}-YX^{\prime}\right)\left(T\dot{M}-M\dot{T}\right)+16\left(Q^{\prime}\dot{Z}-Z^{\prime}\dot{Q}\right)\right)\right].\nonumber 
\end{eqnarray}
Now, it is useful to perform the following decomposition in the asymptotic
region which is compatible with \eqref{eq:A Connections}

\[
G=g\left(t,\phi\right)b\left(r\right),
\]
with

\[
g\left(t,\phi\right)=\exp\left[\sqrt{\frac{8\pi}{\kappa}}\varphi L_{0}+\sqrt{\frac{6\pi}{\kappa}}\psi W_{0}\right],
\]
and where $b\left(r\right)$ is an arbitrary gauge group element depending
on the radial coordinate that generically can be decomposed as 
\[
b\left(r\right)=b_{\left(+\right)}b_{\left(0\right)}b_{\left(-\right)},
\]
with

\begin{eqnarray*}
b_{\left(+\right)} & = & e^{\left(b_{1}L_{1}+\bar{b}_{1}W_{1}+\bar{b}_{2}W_{2}\right)},\qquad b_{\left(0\right)}=e^{\left(b_{0}L_{0}+\bar{b}_{0}W_{0}\right)},\qquad b_{\left(-\right)}=e^{\left(b_{-1}L_{-1}+\bar{b}_{-1}W_{-1}+\bar{b}_{-2}W_{-2}\right)}.
\end{eqnarray*}
Consistency with \eqref{eq:GWZW} then implies the following conditions
\[
\Phi=b_{0}\left(r\right)+\sqrt{\frac{8\pi}{\kappa}}\varphi\left(t,\phi\right),\qquad\Phi_{W}=\bar{b}_{0}\left(r\right)+\sqrt{\frac{6\pi}{\kappa}}\psi\left(t,\phi\right),
\]
\[
X=b_{-1}\left(r\right),\qquad Y=\bar{b}_{-1}\left(r\right),\qquad Z=\bar{b}_{-2}\left(r\right),
\]
\[
Q=e^{-2\sqrt{\frac{8\pi}{\kappa}}\varphi}\bar{b}_{-2}\left(r\right),\qquad M=e^{-\sqrt{\frac{8\pi}{\kappa}}\varphi}\left(\bar{b}_{1}\left(r\right)\cosh\left(2\sqrt{\frac{6\pi}{\kappa}}\psi\right)-b_{1}\left(r\right)\sinh\left(2\sqrt{\frac{6\pi}{\kappa}}\psi\right)\right),
\]
\[
T=e^{-\sqrt{\frac{8\pi}{\kappa}}\varphi}\left(b_{1}\left(r\right)\cosh\left(2\sqrt{\frac{6\pi}{\kappa}}\psi\right)-\bar{b}_{1}\left(r\right)\sinh\left(2\sqrt{\frac{6\pi}{\kappa}}\psi\right)\right).
\]
Note that since, $X$, $Y$ and $Z$ depend only on the radial coordinate,
then the WZ term in eq. \eqref{eq:I2ap} identically vanishes.

\bibliographystyle{JHEP}
\bibliography{review}

\end{document}